\documentclass{aa}  
\usepackage{graphicx}
\usepackage{txfonts}
\usepackage{natbib}
\bibpunct{(}{)}{;}{a}{}{,} 
\usepackage{xcolor}
\begin{document}


   \title{The origin of the stellar mass-size relation of satellite galaxies in the COLIBRE simulations}

   \author{Bruno M. Celiz
          \inst{1,2}\fnmsep\thanks{\email{bruno.celiz@mi.unc.edu.ar}}
           \and
           Julio F. Navarro\inst{3,4}
          \and
          Joop Schaye\inst{5}
          \and
          Mario G. Abadi\inst{2}
          \and
          Alejandro Benitez-Llambay\inst{6}
          \and
          Evgenii Chaikin\inst{5,7}
          \and
          Carlos S. Frenk\inst{7}
          \and
          Filip Huško\inst{5}
          \and
          Aaron D. Ludlow\inst{8}
          \and
          Sylvia Ploeckinger\inst{9}
          \and
          Alexander J. Richings\inst{10,11}
          \and
          Matthieu Schaller\inst{5,12}          
          }

   \institute{Facultad de Matemática, Astronomía, Física y Computación, UNC, Medina Allende s/n, X5000HUA, Córdoba, Argentina
    \and
        Instituto de Astronomía Teórica y Experimental, CONICET--UNC, Laprida 854, X5000BGR, Córdoba, Argentina
     \and
         Department of Physics and Astronomy, University of Victoria, Victoria, BC, V8P 5C2, Canada
     \and
         Crimson Distinguished Professor, KU-KIST Green School, Korea University.
    \and
        Leiden Observatory, Leiden University, PO Box 9513, 2300 RA Leiden, the Netherlands
    \and
        Dipartimento di Fisica G. Occhialini, Universit\`a degli Studi di Milano Bicocca, Piazza della Scienza, 3 I-20126 Milano MI, Italy
    \and
        Institute for Computational Cosmology, Department of Physics, University of Durham, South Road, Durham, DH1 3LE, UK
    \and
        International Centre for Radio Astronomy Research, University of Western Australia, 35 Stirling Highway, Crawley, Western Australia 6009, Australia
    \and
        Department of Astrophysics, University of Vienna, Türkenschanzstrasse 17, A-1180 Vienna, Austria
    \and
        Centre for Data Science, Artificial Intelligence and Modelling, University of Hull, Cottingham Road, Hull, HU6 7RX, UK
    \and
        E. A. Milne Centre for Astrophysics, University of Hull, Cottingham Road, Hull, HU6 7RX, UK
    \and
        Lorentz Institute for Theoretical Physics, Leiden University, PO Box 9506, 2300 RA Leiden, The Netherlands
    }

   \date{Received XXX; accepted YYY}
 
    \abstract 
    {We study the stellar mass-size relation of satellite galaxies in the COLIBRE suite of cosmological hydrodynamical simulations. Satellites deviate from the relation that holds for centrals galaxies, where at the high mass end, $\log (M_*/{\rm M}_\odot) > 10.5$, sizes (defined as the 3D half-mass radius $r_{\rm h,*}$) increase systematically with mass ($r_{\rm h,*} \propto M_*^{0.5}$), whereas at lower masses, $8 < \log(M_*/{\rm M}_\odot) < 10.5$, the relation flattens and galaxy size becomes, on average, almost independent of mass ($r_{\rm h,*} \approx 3$ kpc). At $z=0$, dwarf satellites (defined as those with $8 < \log(M_*/{\rm M}_\odot) < 9$)  are systematically larger than centrals of similar $M_*$. This trend reverses for bright satellites ($9 < \log(M_*/{\rm M}_\odot) < 10.5$), which are typically smaller than centrals of similar mass. We trace these trends to evolutionary processes affecting satellites after infall into the haloes of more massive hosts. At infall, dwarf satellites are typically gas-rich, dark matter-dominated systems with relatively large baryon-induced cores. These satellites quench rapidly after losing their gas to ram pressure, which prompts an immediate impulsive expansion due to the shallowing central potential, followed by secular expansion as their cored dark matter haloes are gradually stripped by tides. In contrast, the inner regions of bright satellites are baryon-dominated and resilient to tides. Centrally concentrated star formation increases their stellar mass, leading to smaller sizes and higher stellar metallicities (by $\approx 0.2$ dex) than those of centrals of similar mass. These distinct satellite evolutionary pathways lead to identifiable features in the mass-size-metallicity relations that may be compared with observations.}


   \maketitle

\section{Introduction} \label{sec:intro}

Satellite galaxies serve as a testbed of the $\Lambda$ Cold Dark Matter ($\Lambda$CDM) paradigm, which predicts that dark-matter (DM) haloes are populated by a hierarchy of subhaloes accreted throughout cosmic time \citep{Frenk1988,Navarro1997}. In haloes, the co-evolution of baryonic and dark matter regulates the assembly of galaxies and sets their fundamental properties, such as the mass and size of the stellar component \citep{ReesOstriker1977,FallEfstathiou1980,Mo1998, WechslerTinker2018,Ferrero2021}. 

Because stars behave as collisionless particles on galactic scales, their spatial and kinematic distributions reflect both the properties of the gas from which they formed and the cumulative gravitational interactions experienced throughout their history. Given the wealth of galaxy evolution information encoded in this phase-space distribution, the Stellar Mass-Size Relation (SMSR) serves as a crucial scaling relation to access it; it provides a vital diagnostic by encoding not only the efficiency of star formation within a given halo but also the physical distribution of where that star formation occurs, as well as environmental effects.

Low-mass galaxies represent a critical frontier for cosmological models, as their shallow potential wells render them exceptionally sensitive to both internal baryonic feedback and external environmental effects \citep{Governato2010, Governato2012,DiCintio2014,Christensen2016,Sawala2016dwarfs}. A number of persistent ``tensions'' with observations have been identified in  $\Lambda$CDM on the scale of dwarf galaxies \citep[see e.g.][for a review]{Sales2022}, in particular regarding their innermost  mass distribution and scaling relations  \citep[][]{Walker2009,Oh2011,Oman2015,Cruz2025}.

For satellite galaxies, a major driver of galactic transformation is their environment, where non-linear processes reshape their DM, stellar, and gaseous components \citep[see e.g.][]{Kauffmann2004,Wetzel2013,Peng2015, Buck2019}. The internal properties of satellites are thus expected to differ from those of field galaxies of comparable stellar mass if it is preserved \citep[see e.g.][]{Peng2012}. Key mechanisms driving this evolution include tidal stripping and heating \citep{Merritt1983, Gnedin1999}; ram-pressure stripping of gas through hydrodynamic interactions with the circumgalactic or intra-group/cluster medium \citep{GunnGott1972, Abadi1999,McCarthy2008}; and intra-group effects such as encounters with neighbouring galaxies \citep[see e.g., gas starvation and galaxy harassment][]{Larson1980,Moore1996Natur} and the gradual orbital energy loss due to dynamical friction \citep[][]{Chandrasekhar1943,Moore1998, TaylorBabul2004}.

One would expect these processes to significantly alter the scaling relations of satellite galaxies relative to field galaxies, although a detailed accounting of the interplay between them remains elusive. In this context, the satellite SMSR provides a valuable diagnostic of such interplay.
Observationally, comparisons of the SMSR of satellites and centrals across major surveys, such as the Sloan Digital Sky Survey \citep[SDSS, e.g.][]{Wang2020,Rodriguez2021,Abdullah2026} and Satellites Around Galactic Analogs \citep[SAGA, e.g.][]{Asali2025}, indicate that the SMSR of satellites slightly deviate from that of centrals. In particular, \citet{Asali2025} report that the half-light sizes of SAGA satellites are systematically larger than those of isolated galaxies, though only by $\approx 0.1$ dex.

On the simulation side, several studies have used large-scale cosmological simulations to explore the environmental origins of these satellite scaling relations, with mixed results.  For instance, \cite{Carleton2019} argued for an environmental origin of extended satellites, which results from the tidal stripping and heating of cored haloes with masses $10^{10-11} \, {\rm M_{\odot}}$. Their semi-analytic model, applied to DM-only (DMO) simulations of the Illustris suite \citep{Vogelsberger2014}, successfully reproduced observed Ultra Diffuse Galaxy (UDG) sizes and masses in clusters. However, their model could not fully account for systems with baryon-dominated centres or those undergoing extreme mass loss. \cite{Sales2020} and \cite{Benavides2023} utilized the TNG100-1 \citep{Nelson2018} and TNG50-1 simulations \citep{Nelson2019TNG50,Pillepich2019} to argue that at least some UDGs should inhabit high-spin haloes, where they may form irrespective of the presence of a DM core.

In addition, other authors have used high-resolution zoom-in simulations to analyse the satellite populations of Milky Way (MW) analogues. For example, \citet[][using APOSTLE]{Fattahi2018} and \citet[][using NIHAO]{Buck2019} found that satellites of MW-like galaxies generally trace the same scaling relations as field dwarfs. Both studies reported only subtle environmental signatures, such as a mild evolution in the mass-metallicity relation \citep[c.f.][]{Bahe2017} and a minor reduction in galaxy size corresponding to stellar mass loss.

We address the structural evolution of low-mass galaxies upon infall using the latest suite of cosmological hydrodynamical simulations, COLIBRE \citep[COLd Ism and Better REsolution,][]{Schaye2025,Chaikin2026}, which faithfully reproduce galaxy properties across cosmic time, such as the stellar mass function and star formation rates \citep{Chaikin2025}, atomic and molecular mass fractions \citep{Schaye2025}, the Kennicutt-Schmidt relation \citep{Lagos2026}, and the gas-phase mass-metallicity relation \citep{Sharda2026}. Furthermore, its sophisticated treatment of the multi-phase interstellar medium and relatively high numerical resolution are able to resolve the internal structural changes of galaxies in different environments, down to the dwarf galaxy regime $(M_* \lesssim 10^9 \, {\rm M_{\odot}})$.

Most relevant to this work, \cite{Ludlow2026} used the COLIBRE simulations to study the size–mass and angular momentum–mass relations of central galaxies in different COLIBRE simulations. In addition, \cite{Forouhar2026} analysed the morphologies of COLIBRE galaxies at $z=0$, and reported that galaxy morphology depends mainly on stellar and gas mass rather than halo mass, and on whether a galaxy is a ``central'' (i.e., the most massive galaxy of its immediate environment) or a ``satellite'' system orbiting a more massive host. \cite{He2026} have studied the dynamics of COLIBRE satellites and the co-evolution of their stellar and DM mass, and report that stellar tidal stripping becomes significant only after a subhalo loses more than $\approx$95\% of its originally bound mass after infalling into a more massive system.

This paper is organised as follows: Section \ref{subsec:simu} describes the COLIBRE simulations, while Section \ref{subsec:glx_sample} outlines our  definition of galaxy samples. Section \ref{SecResults} presents our main results, including (i) the SMSR of central and satellite galaxies at $z=0$ (Section \ref{subsec:SMSR_results}); (ii) the relation between satellite mass, cosmic infall time and satellite survivor fraction (Section \ref{subsec:InfallTimeSats}); (iii) how satellites evolve in the SMSR plane after infall and their mass density profiles (Section \ref{subsec:SMSR_TemporalEvolution}); and (iv) the typical metallicity and DM-fraction of these satellites (Section \ref{subsec:SatPropsObs}). We summarise our findings in Section \ref{SecConc}.

\section{Methods} \label{SecMethods}

\subsection{Simulation} \label{subsec:simu}

For this study we used the COLIBRE suite of cosmological hydrodynamical simulations \citep{Schaye2025,Chaikin2026} of a  $\Lambda$CDM Universe consistent with the Dark Energy Survey year three $\Lambda$CDM cosmology \citep[DES Y3 ‘3×2pt + All Ext.’,][$h = 0.681;~\Omega_{\rm m} = 0.306;~\Omega_{\rm b} = 0.0486;~\sigma_8 = 0.807$]{Abbott2022}. 

COLIBRE improves upon other state-of-the-art cosmological simulations in several key respects: (i) it super-samples the DM using four times as many DM particles as baryonic particles in the initial conditions, so that all particle species have similar masses; the larger number of DM particles per halo increases the relaxation time, together suppressing the spurious transfer of energy from DM to stellar particles \citep[see][]{SteinmetzWhite1997,Revaz2018,Ludlow2019,Ludlow2021,Ludlow2023}; (ii) it uses a highly-parallel astrophysical simulation code built on a task-based parallelisation scheme \citep[SWIFT,][]{Schaller2024}, and (iii) a new structure and substructure finder based on tracking the evolution of bound structures across snapshots \citep[HBT-HERONS,][]{Forouhar2025}. 

All sub-grid prescriptions are new or have been significantly modified from the versions adopted for the precursor EAGLE \citep[Evolution and Assembly of GaLaxies and their Environments][]{Schaye2015,Crain2015}. In particular, (i) gas can cool below $10^4$ K, and no equation of state is imposed on high-density gas \citep{Nobels2024}; (ii) the growth and composition of dust grains are simulated in a self-consistent way \citep{Correa2026,Trayford2026}; self-shielding and molecules are included and coupled to the dust physics \citep{Ploeckinger2025}; (iii) pre-supernova stellar feedback is included \citep{BL2026}; and (iv) the numerical prescriptions for supernova and Active Galactic Nuclei (AGN) sampling and feedback are improved \citep{Chaikin2023,Husko2026}. Four parameters of the feedback model (three for supernova feedback and one for AGN feedback) were optimised at low resolution using Gaussian process emulators to match two fundamental scaling relations of the galaxy population at $z=0$: the galaxy stellar mass function and the mass-size relation in the stellar mass range $10^9 < M_*/{\rm M}_{\odot} < 10^{11.3}$ \citep{Chaikin2026}. At higher resolutions, this calibration was performed manually, using the best-fit parameters inferred at the low resolution as the initial guess.

Together, DM super-sampling (which prevents artificial two-body heating of stars) and low-temperature gas cooling enable COLIBRE to reliably model sub-kiloparsec stellar distributions and ISM morphology down to the dwarf galaxy regime \citep[see e.g.][]{Zeng2024,Celiz2025}

The COLIBRE suite is also well suited for convergence tests, including simulations with different volumes, resolutions, and AGN feedback model. L400m7 and L200m6 are the two flagship simulations that have already reached $z=0$, following $5 \times 3008^3$ particles in cubic volumes of 400 and 200 cMpc on a side and yielding particle masses, for both baryons and DM, of $\sim$$10^7$ and $\sim$$10^6 \, {\rm M_{\odot}}$, respectively. There is also a higher-resolution model that follows $5 \times 752^3$ particles of mass $\sim$$10^5 \, {\rm M_{\odot}}$ in a 25 cMpc box (``L025m5''). 

All simulations start from initial conditions set at $z=63$, and the state of each simulation is saved in 128 snapshots between $z=30$ and 0 at intervals of expansion factor $\Delta a \approx 0.01 \, a$. Each hydrodynamic simulation was also performed in a DMO mode, utilising the same particles and Fourier modes in the initial conditions, but evolving them with gravity only. Additionally, each hydrodynamic run was performed twice, using either the Thermal or Hybrid AGN feedback model \citep{Husko2026}. In Appendix \ref{AppNumConv} we show how the median SMSR for centrals and satellites changes with particle resolution and volume, comparing the L025m5, L025m6 and L025m7 runs for the $25^3 \, {\rm Mpc^3}$ volume, and the L200m6 and L200m7 runs for the $200^3 \, {\rm Mpc^3}$ volume.

Of all available runs, the fiducial Thermal\_L200m6 run (L200m6 hereafter) is the best suited for our study because it offers the best trade-off between particle resolution and volume. In L200m6, DM particles have a mass of m$_{\rm{DM}} = 2.42 \times 10^6 \, {\rm M_{\odot}}$ and baryons have an initial mass of m$_{\rm{baryon}} = 1.84 \times 10^6 \, {\rm M_{\odot}}$. The Plummer-equivalent gravitational softening for all particles varies with redshift ($\epsilon_{\rm{DM,*}} \approx 1.8/(1+z)$ pkpc), and remains fixed at $0.7$ pkpc at $z \leq 1.57$.

\begin{figure}
    \includegraphics[width=\columnwidth]{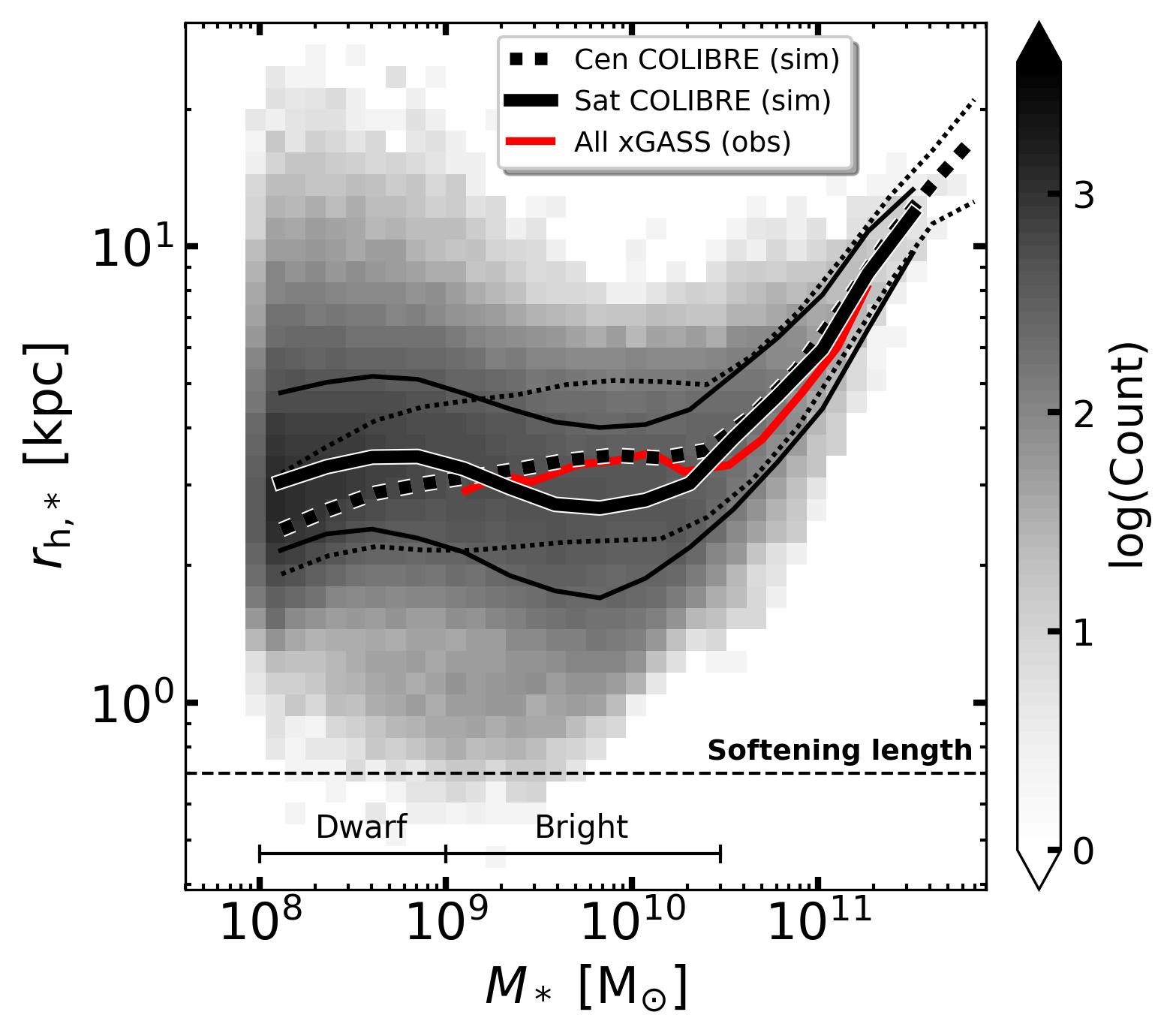}

    \caption{Stellar half-mass radius ($r_{h,*}$, 3D) as a function of stellar mass ($M_{*}$) for satellite galaxies at $z = 0$ in the COLIBRE L200m6 simulation. The grey scale indicates the logarithm of the number of satellites in each bin, as shown in the side bar. The black solid line shows the median stellar mass-size relation of satellites, whereas that of centrals is shown by the thick black dotted line. The 16th and 84th percentiles are shown as thin lines. We show the results for observed  galaxies reported by \citet[][]{Hardwick2022} using xGASS (solid red line), converted to 3D stellar half-mass radii using $r_{\rm h,*}=4/3\,R_{\rm h,*}$. At the massive end ($\log (M_{*}/{\rm M}_{\odot}) > 10.5$)  COLIBRE galaxies follow roughly the scaling relation $r_{\rm h,*} \propto M_*^{0.5}$, but less massive galaxies do not. For centrals, the trend flattens at lower masses, but for satellites the trend is non-monotonic. ``Bright'' satellites, with $9 \lesssim \log (M_{*}/{\rm M}_{\odot}) < 10.5$, are typically smaller than centrals with the same stellar mass, but dwarf satellites ($\log (M_{*}/{\rm M}_{\odot}) \lesssim 9$) are typically larger than centrals of similar mass.}

    \label{SMSR_sats_vs_cent}
\end{figure}

\subsection{Simulated galaxies}
\label{subsec:glx_sample}

We identify galaxies using the catalogues generated by two structure-finding algorithms: DM haloes are identified using a friends-of-friends \citep[FoF,][]{PressDavis1982}
halo finder applied to the DM particles, with a linking length of 0.2 times the mean interparticle separation (baryonic particles are then linked to the nearest DM particle that belongs to a FoF halo and is within the linking length); then a subhalo finder, HBT-HERONS \citep{Han2018,Forouhar2025}, is used to identify self-bound subhaloes. HBT-HERONS is a modified version of the hierarchical bound-tracing algorithm HBT+ \citep{Han2018}, which analyses consecutive outputs of the simulation, from early to late times, and uses a set of particles that belong to the main progenitor to to robustly track self-bound subhaloes with at least 20 bound particles and 10 tracer (i.e. DM or stellar) particles. Fundamentally, it minimizes common artifacts of structure finders, such as central-galaxy swapping during major mergers or temporarily losing substructures across snapshots.

Finally, a large number of halo and subhalo properties are computed using the Spherical Overdensity and Aperture Processor tool \citep[SOAP,][]{McGibbon2025}. These properties can be measured accounting for all or only bound particles, across different 3D and projected spatial apertures.

We restrict our analysis to galaxies in the stellar mass\footnote{Throughout this paper, $\log$ denotes the base-10 logarithm, $\log_{10}$.} range $\log (M_{*}/{\rm M}_{\odot}) > 8$. For central galaxies (defined as that with the highest mass and lowest specific orbital kinetic energy within a FoF group), we also impose a lower bound on virial mass\footnote{In the present work, virial quantities are defined at the radius enclosing 200 times the critical density for closure.}, $\log (M_{\rm 200}/{\rm M}_{\odot}) > 10$. This cut removes backsplash galaxies while ensuring sample completeness across the stellar mass range of interest. We measure galaxy properties using all bound particles within 50 pkpc centred on the most bound particle of the subhalo. This choice follows \cite{Schaye2025} for central galaxies, motivated by the finding of \cite{deGraaff2022} that this aperture reproduces masses inferred from Sérsic fits to mock Sloan Digital Sky Survey images of galaxies in the EAGLE simulation. For satellite galaxies, considering only self-bound particles prevents contamination from the surrounding host halo. Because a satellite’s bound mass is concentrated near its centre, the exact aperture radius has no practical impact on the derived quantities, and 50 kpc comfortably encompasses all bound stellar particles.

To study the SMSR of simulated galaxies, we use the 3D stellar half-mass radius $r_{\rm h,*}$, and the stellar mass $M_*$, defined as the sum of the masses of all stellar particles bound to the subhalo within the aforementioned aperture. We verified that our results remain consistent when using 2D projected quantities instead of 3D measurements. Thus, we prioritize 3D properties over 2D projections to capture intrinsic physical structures without projection biases.

In total, we end up selecting $158{,}154$ central galaxies and $131{,}174$ satellites at redshift $z=0$ in the L200m6 simulation. The lowest mass satellites, with $\log (M_{*}/{\rm M}_{\odot}) \approx 8$, are constituted by at least 50 stellar particles.

\section{Results} \label{SecResults}

\begin{figure}
    \includegraphics[width=\columnwidth]{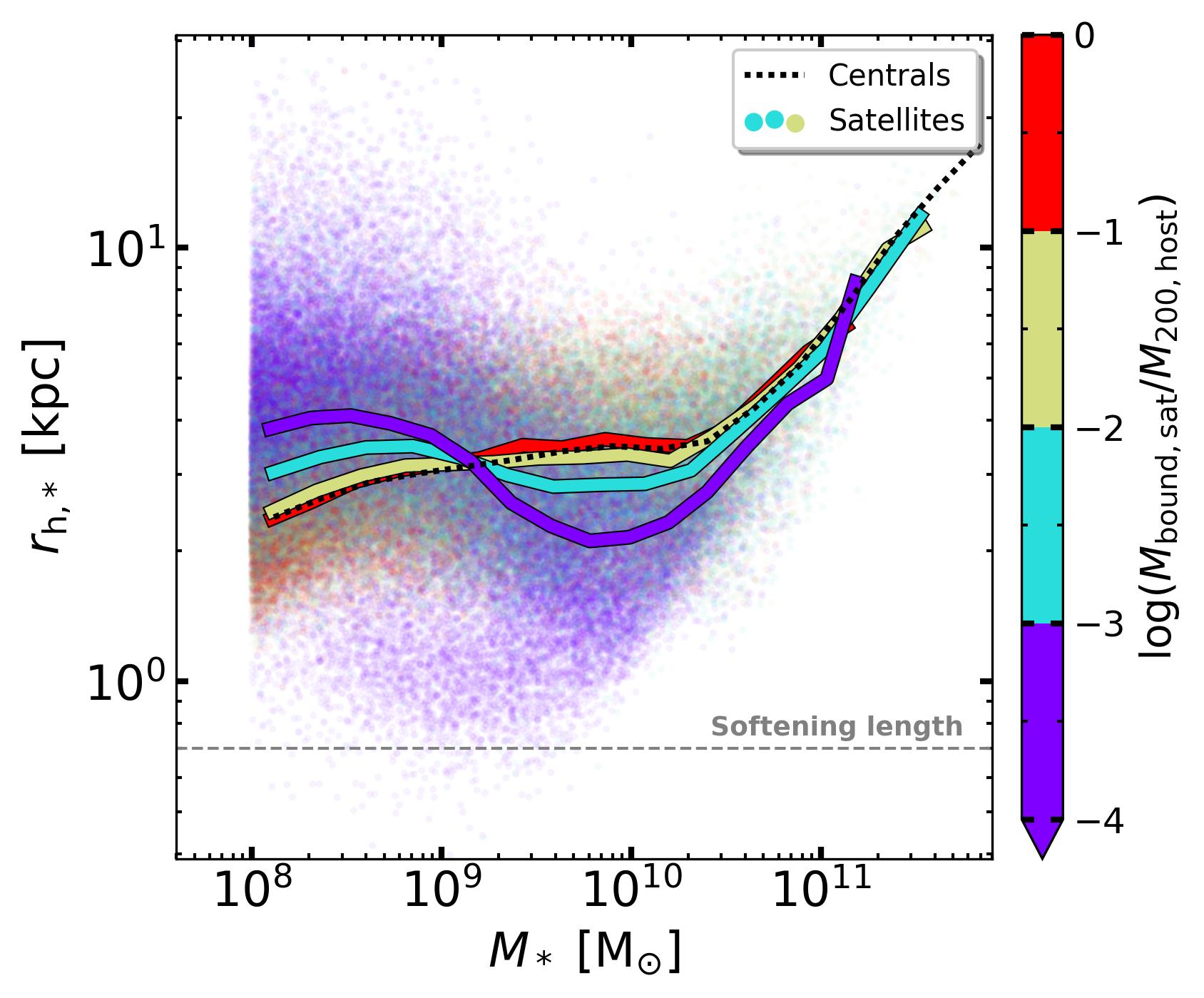}

    \caption{Stellar half-mass radius ($r_{h,*}$, 3D) as a function of stellar mass ($M_{*}$) for simulated satellites at $z=0$, shown as dots coloured by their total bound mass relative to the virial mass of the halo that they inhabit. We compare the median trend of subsamples of satellites with similar mass ratio (coloured solid lines) to the median trend of central galaxies (black dotted line). Satellites with similar total mass to that of their host (red) exhibit similar sizes than centrals at each mass. However, satellites that are progressively less massive than their hosts (lower mass ratios), systematically deviate from the SMSR of centrals. In those cases, dwarf satellites are typically larger, and bright galaxies (up to $\log (M_{*}/{\rm M}_{\odot}) \sim 10.5$) are typically smaller than centrals of similar stellar mass.}
    
    \label{SMSR_sats_COLIBRE}
\end{figure}

\subsection{Galaxy stellar mass-size relation}
\label{subsec:SMSR_results}

The main goal of this study is to understand how the SMSR of satellite galaxies differs from that of central galaxies in  COLIBRE. We explore this in Fig. \ref{SMSR_sats_vs_cent}, where the median and 16th-84th percentiles (black solid lines) of the satellite SMSR are compared with those of centrals (black dotted lines). For comparison, we also show in red the SMSR reported by \citet[][]{Hardwick2022} using the extended GALEX Arecibo SDSS Survey (xGASS), as shown in \citet[][]{Ludlow2026}, where 2D radii ($R_{\rm h,*}$) have been transformed to 3D assuming $r_{\rm h,*}=4/3\,R_{\rm h,*}$ \citep[][]{Hernquist1990,Wolf2010}. As mentioned in Sec. \ref{subsec:simu}, COLIBRE was calibrated to match these data \citep{Chaikin2026}.

Central galaxies follow a median trend of monotonically increasing size with mass, with a rather mild increase from $\sim$2 kpc to $\sim$4 kpc in the mass range $8 < \log (M_{*}/{\rm M}_{\odot}) < 10.5$, and steepens for more massive galaxies. Satellites, on the other hand, exhibit a non-monotonic SMSR that ``wiggles'' around the SMSR of centrals: ``dwarf satellites'' ($8 < \log (M_{*}/{\rm M}_{\odot}) < 9$) are typically larger than centrals of similar mass, whereas  ``bright satellites'' ($9 < \log (M_{*}/{\rm M}_{\odot}) < 10.5$) are typically smaller than centrals. For the most massive satellites ($\log (M_{*}/{\rm M}_{\odot}) > 10.5$), the trend steepens and closely follows that of centrals.

To investigate the origin of the differences between the SMSR of satellites and centrals, we decompose the satellites into subsamples based on the mass ratio between a satellite and its host halo, $M_{\rm bound,sat}/M_{\rm 200,host}$. Here, $M_{\rm bound,sat}$ is the total mass bound to the satellite (within 50 pkpc of its centre), and $M_{\rm 200,host}$ is the virial mass of the host galaxy. The median SMSR for subsamples of galaxies with different values of the mass ratio are shown with coloured solid lines in Fig. \ref{SMSR_sats_COLIBRE}. 

The differences between satellite and central SMSRs are clearly exacerbated for the least massive satellites,  relative to their host. Satellites with high mass ratios ($\log(M_{\rm bound,sat}/M_{\rm 200,host}) > -1$, shown in red) closely follow the SMSR of centrals, indicating that they have been largely unaffected by the host environment. Conversely, galaxies with progressively lower mass ratios deviate gradually from the centrals' trend. Those in the lowest mass-ratio bins ($\log(M_{\rm bound,sat}/M_{\rm 200,host}) < -3$, shown in purple) exhibit the most pronounced SMSR deviations.

Studies using major surveys such as SDSS \citep[e.g.][]{Wang2020,Rodriguez2021,Abdullah2026} and SAGA \citep[e.g.][]{Asali2025}, find that the SMSR of satellite galaxies barely deviates from that of centrals, unlike our findings. In light of our Figure 2, this discrepancy may stem from selection effects; observations are naturally biased toward more massive satellites \citep[see e.g.][]{Font2021,Font2022}. 

In COLIBRE, satellites with the lowest mass ratios ($\log(M_{\rm bound,sat}/M_{\rm 200,host}) < -3$) represent 36.2\% of all $z=0$ satellites, and drive the systematic differences between the satellite and centrals' SMSR noted in Fig. \ref{SMSR_sats_vs_cent}. In the following sections, we focus on this extreme population, where the systematic differences between centrals and satellites are most pronounced.

\begin{figure}
    \centering
    \includegraphics[width=0.87\columnwidth]{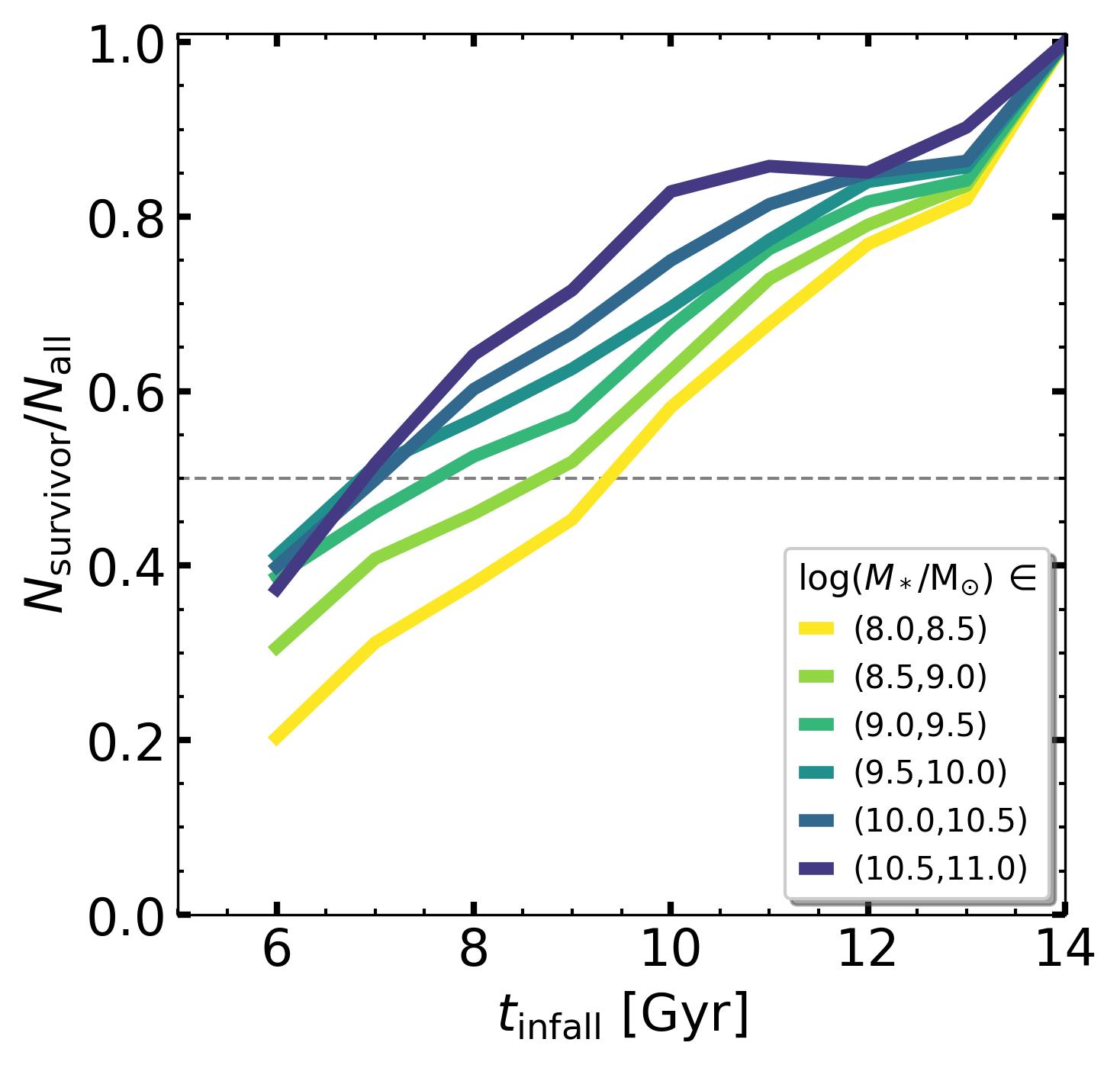}
    
    \caption{Fraction of  satellites that survive  to $z=0$ as a function of their infall time, split in bins of stellar mass at infall (colour lines). The surviving fraction increases with accretion time and mass; few early accreting satellites survive to the present, especially low-mass ones.}
    \label{f_Survivors_t_infall}
\end{figure}

\begin{figure*}
    \centering
    \includegraphics[width=2\columnwidth]{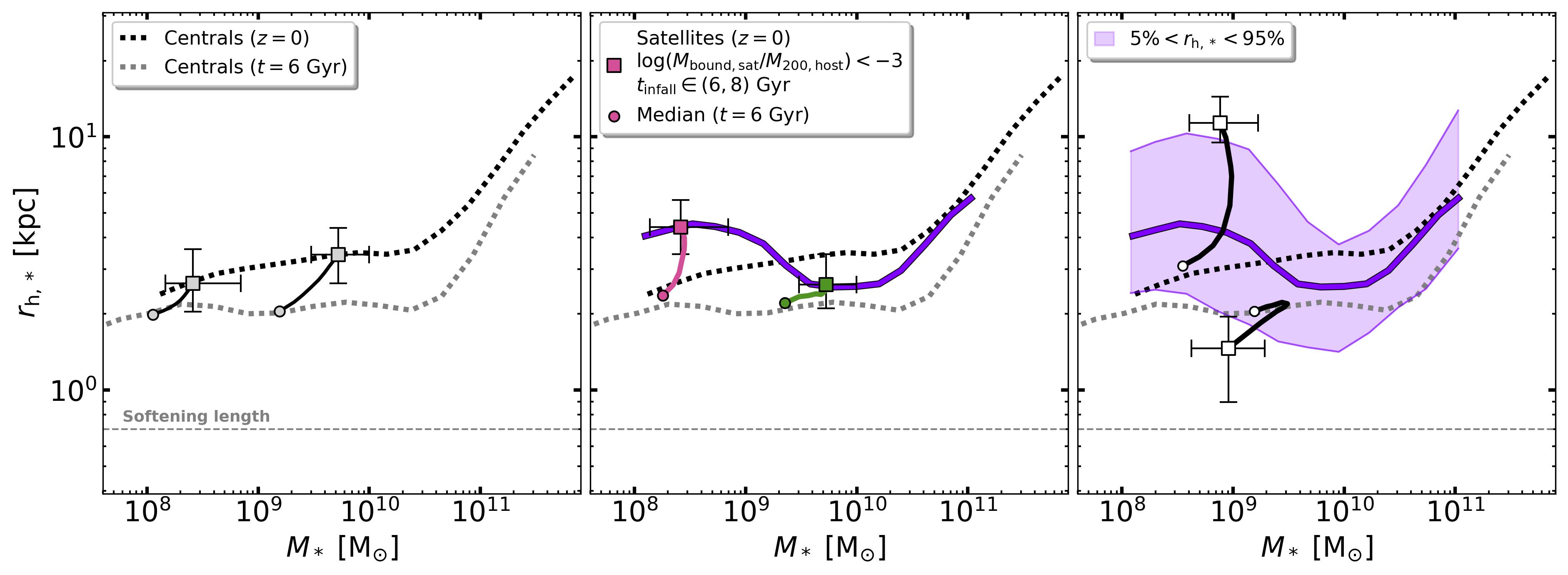}
    
    \caption{Median evolutionary trajectories of COLIBRE galaxies in the mass-size plane, from cosmic time $t=6$ Gyr (circles) to $z=0$ (squares).
    Left panel: Evolutionary tracks for two central galaxy subsamples, drawn from two logarithmic bins of $z=0$ stellar mass ($[8, 8.8]$ and $[9.5,10]$, in solar units). The median trend for central galaxies at $t = 6$ Gyr (grey dotted line) and that of central galaxies at $z=0$ (black dotted line) are shown for reference.
    Middle panel: median evolution of two satellite subsamples with mass ratios $\log(M_{\rm bound,sat}/M_{\rm 200,host}) < -3$ (purple solid line, as in Fig. \ref{SMSR_sats_COLIBRE}), tracked backwards from $z=0$ to cosmic time $t = 6$ Gyr (pre-infall). Coloured squares and error bars indicate the regions of the SMSR plane encompassed by each subsample in all three panels. Tracks represent the ``dwarf'' (pink) and ``bright'' (dark green) galaxy regimes. Galaxies are chosen to have infall times $6 < t_{\rm infall}/{\rm Gyr} < 8$. Bright satellites grow more in mass but less in size compared to dwarfs, which grow in size but not much in mass.
    Right panel: Trajectories for extreme dwarf cases selecting satellite galaxies with $z=0$ sizes below the 5th percentile and above the 95th percentile in the stellar mass range $8.5 < \log (M_{*}/{\rm M}_{\odot}) < 9.5$. While compact dwarfs ($<$5th percentile) typically originate as more massive systems that subsequently undergo significant mass loss, very extended dwarfs ($>$95th percentile) were dwarf galaxies with sizes above the median before infall.}
    \label{SMSR_tracks_subsamples}
\end{figure*}

\subsection{Infall time and satellite survival}
\label{subsec:InfallTimeSats}

Our satellite sample consists of all systems bound to a more massive host at $z=0$. Recent work has shown that this may constitute a rather biased sample, where many satellites on small pericentric orbits or those with short orbital times are missing because of the effects of limited numerical resolution  \citep[see e.g.][]{vandenBosch2018,ErraniNavarro2021,Errani23,SantosSantos2025}.

The work of \citet{Errani23}, in particular, indicates that the tidal fate of satellite systems is determined by the structure of their DM (sub)haloes. Cuspy ``NFW-like'' DM subhaloes \citep{Navarro1996a,Navarro1997} are strongly resilient to tides and should always leave behind a bound remnant. On the other hand, should subhaloes develop a constant-density ``core'', perhaps via baryonic outflows \citep[see e.g.][]{Navarro1996b, PontzenGovernato2012, Read2016}, they become vulnerable to full tidal disruption, depending on the size of the core and on the strength of the tidal field they experience. 
Limited numerical resolution (i.e., small particle number and gravitational softening) also results in the formation of central ``cores'', which may artificially lead to the full tidal disruption of a substantial fraction of the satellite population. This artificial disruption is expected to affect especially the earliest accreting, most poorly-resolved (low-mass) satellites.  

A simple metric for these effects is the fraction of satellites that have survived self-bound to $z=0$. Fig. \ref{f_Survivors_t_infall} shows this ``survivor'' fraction as a function of infall time, binned by stellar mass (measured at infall). Here, infall time ($t_{\rm infall}$) is defined as the cosmic time at which a central galaxy is accreted into a more massive system and becomes a satellite. This figure shows that satellite survival depends critically on how early a galaxy was accreted by its host and on its mass \citep[see also][]{Yun2019}. Dwarf satellites are more severely affected than bright satellites; only $34\%$ of dwarfs accreted between $6 < t_{\rm infall}/{\rm Gyr} < 8$ survive to the present day, compared with $50\%$ of bright satellites. Therefore, the dwarf satellite population identified at $z=0$ is heavily biased towards systems that have become satellites relatively recently. In what follows, we account for this bias by analysing our results within subsamples matched in infall time and satellite-to-host mass ratio.

\begin{figure}
    \centering
    \includegraphics[width=0.85\columnwidth]{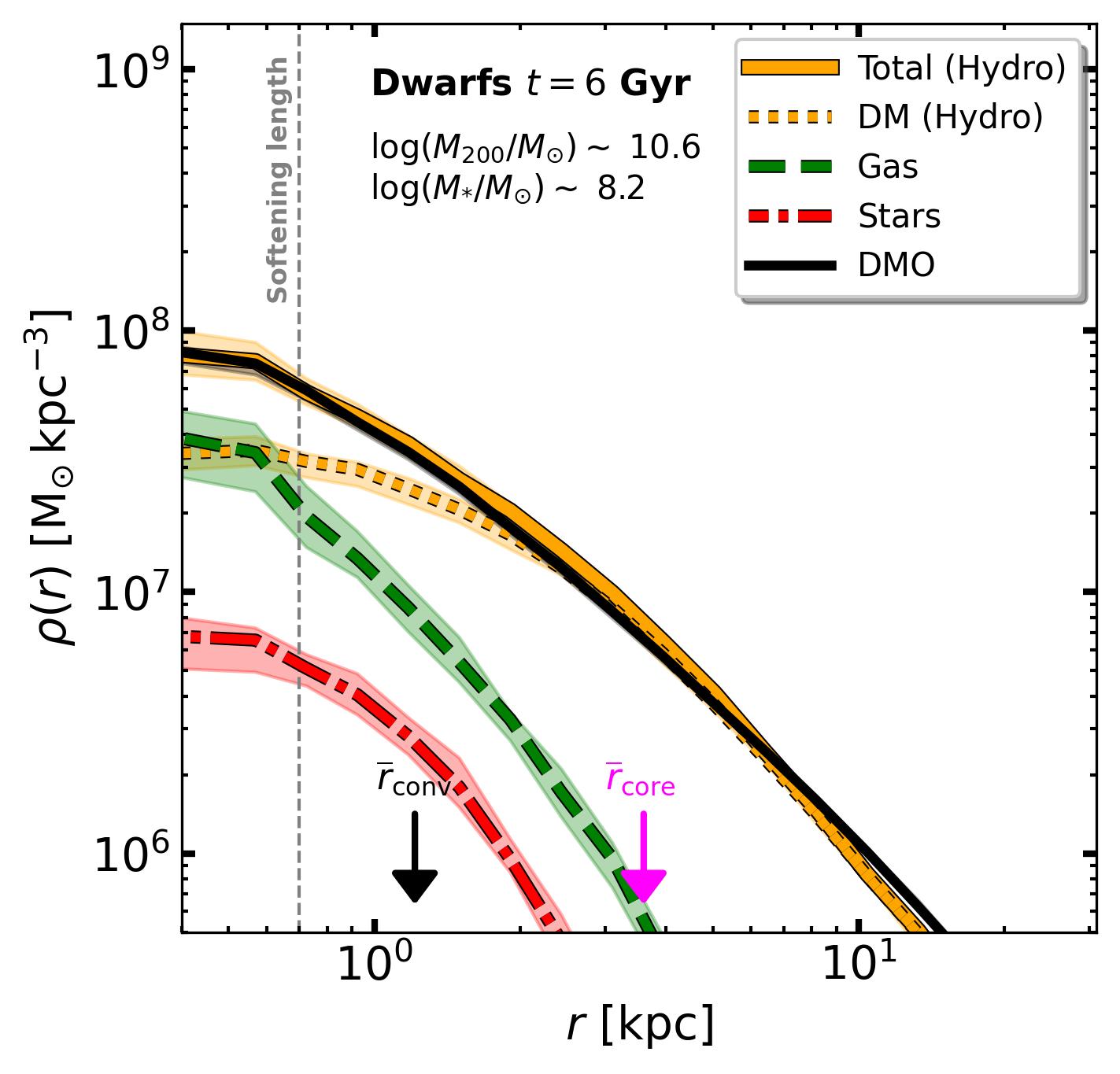}
    \includegraphics[width=0.85\columnwidth]{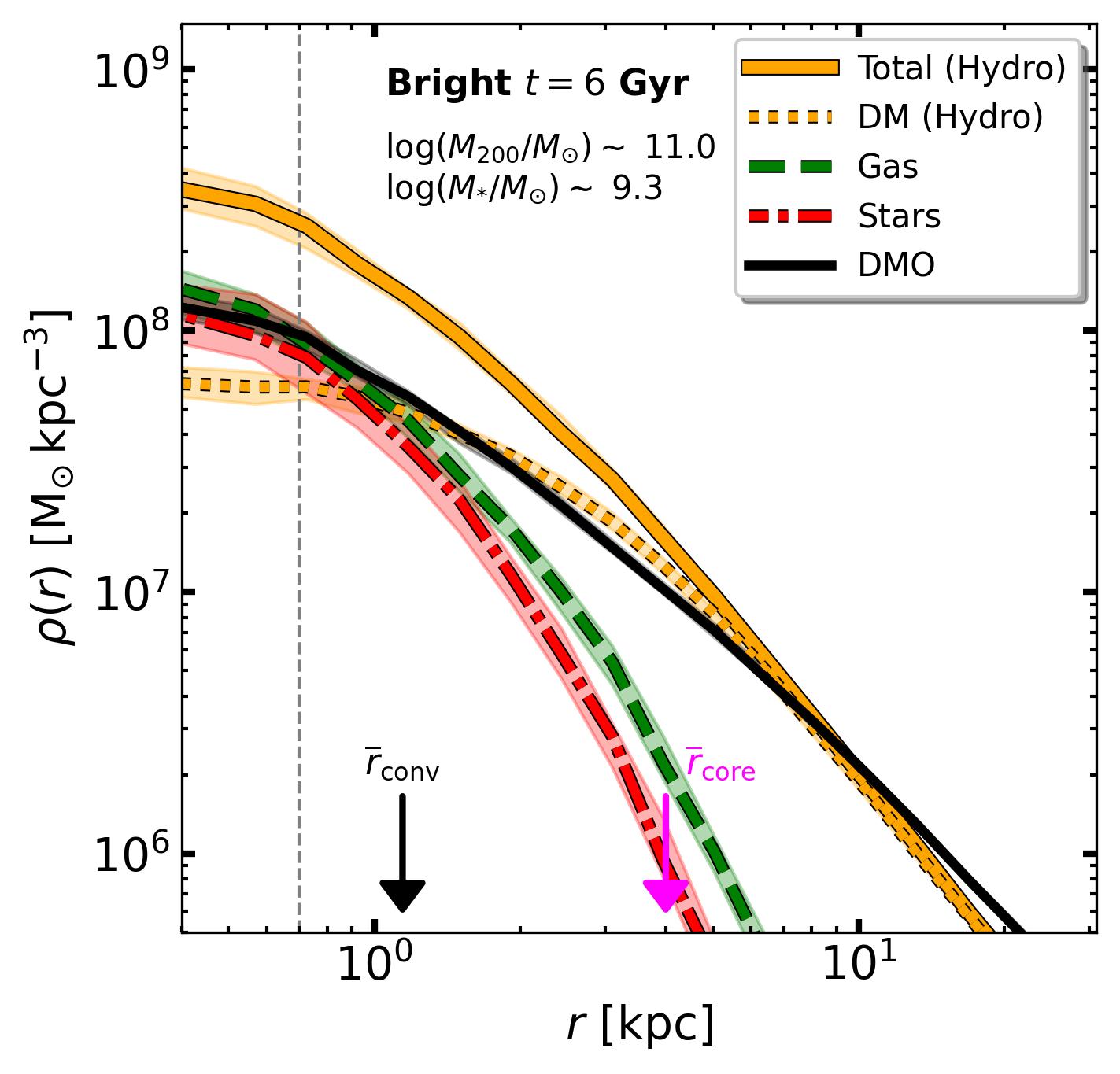}
    
    \caption{Median mass density profiles of the dwarf (top panel) and bright (bottom panel) subsample of satellite galaxies, at cosmic time $t = 6$ Gyr (prior to infall). For comparison, we show the profiles of subhaloes with similar virial mass from the DMO run (black solid line). For the hydrodynamic run (``Hydro''), we show the total mass density profile with an orange solid line. In the hydrodynamic run, the DM component (orange dotted line) exhibits a flatter density profile than in the DMO subhaloes inside $r \lesssim 2$ kpc. In the dwarf regime, the stellar component (red dash-dotted line) is significantly less massive than the gaseous component (green dashed line), while in the bright regime it typically dominates the innermost region, resulting in cuspier total density profiles. Shaded regions encompass the 25-75th percentiles. Black and magenta arrows indicate the median convergence radius for DMO subhaloes \citep[$\bar{r}_{\rm conv}$, as defined in][]{Ludlow2019} and the median DM core radius \citep[$\bar{r}_{\rm core}$, obtained fitting a Burkert profile,][]{Burkert1995}, respectively.}
    \label{stack_hydro_vs_DMO}
\end{figure}

\begin{figure}
    \centering
    \includegraphics[width=0.95\columnwidth]{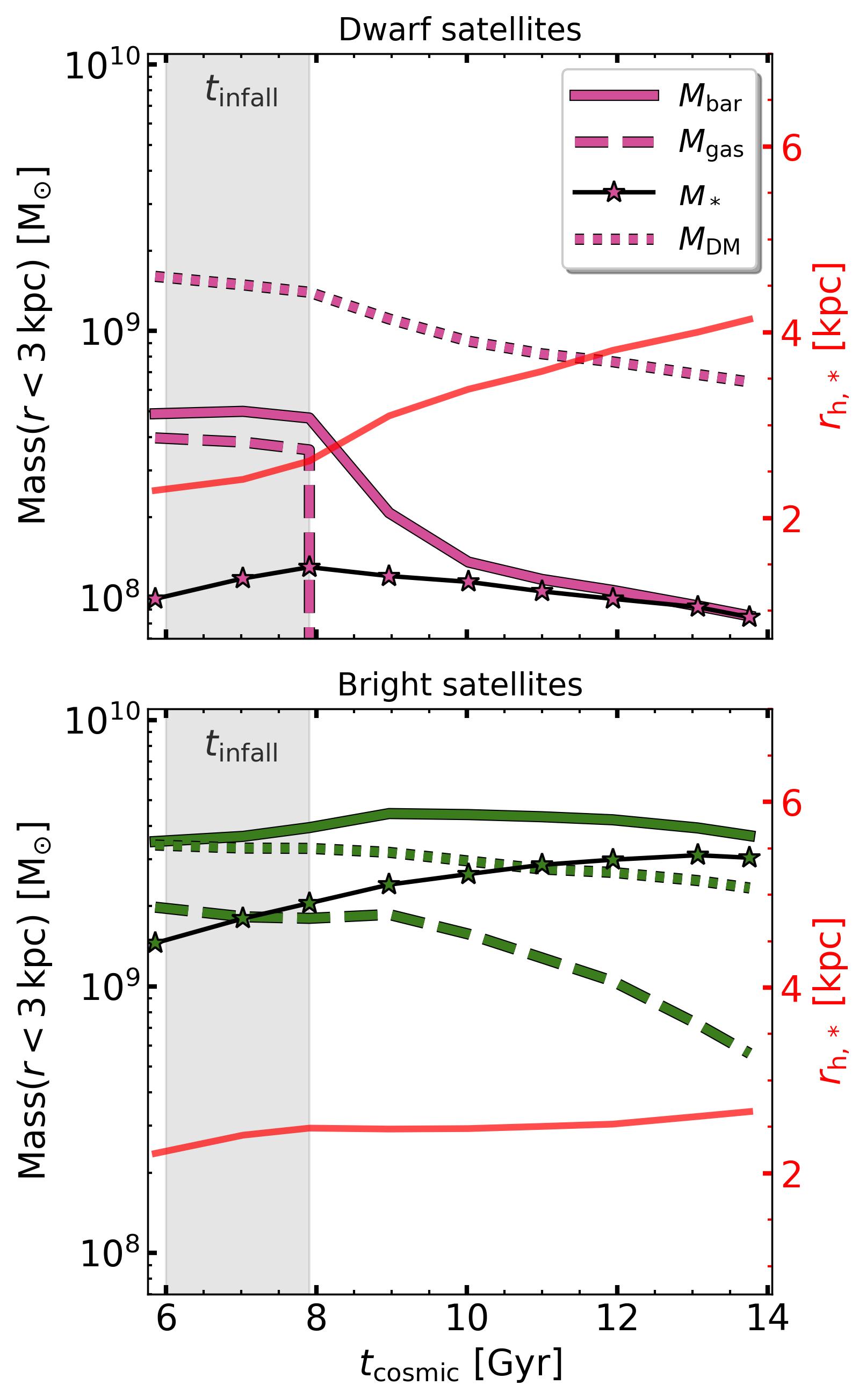}
    
    \caption{Temporal evolution of median mass (coloured lines) of the dwarf (top panel) and bright (bottom panel) subsample of satellites, using all bound particles within an inner aperture of 3 kpc. The median stellar half-mass radius evolution is shown with solid red lines in each panel. These galaxies were selected to have similar infall times ($t_{\rm infall} =$ 6-8 Gyr, grey shaded regions) and very low satellite-host mass ratios ($\log(M_{\rm bound,sat}/M_{\rm 200,host}) < -3$). Dwarf satellites rapidly lose their gas (dashed lines) within 1 Gyr after infall, driving an impulsive stellar expansion due to the shallowing central potential; this is followed by secular tidal expansion as the cored DM halo is gradually stripped. In contrast, bright satellites retain their inner baryon content (solid lines), convert gas into stars (coloured symbols connected by a black line), and maintain a nearly constant size.}
    \label{evo_temp_subsamples_props}
\end{figure}

\begin{figure}
    \centering
    \includegraphics[width=\columnwidth]{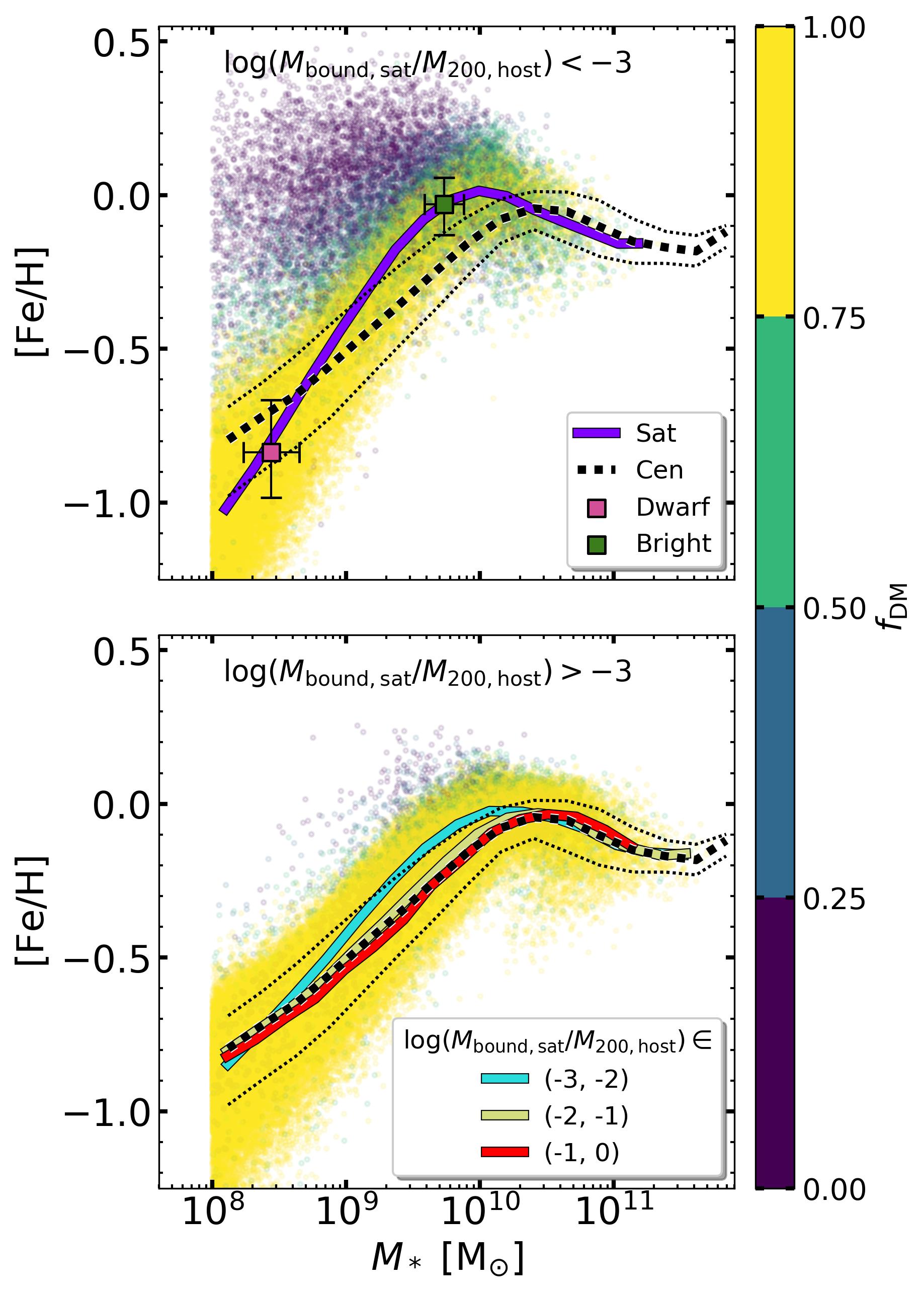}
    
    \caption{Mass-weighted iron-to-hydrogen stellar abundance in solar units (${\rm [Fe/H]}$) as a function of stellar mass for $z=0$ satellites. Each galaxy is coloured by its total DM mass fraction ($f_{\rm DM}$, computed using all bound particles within 50 kpc of each satellite), with darker colours indicating lower values. The top panel shows satellites with the lowest satellite-to-host mass ratios ($\log(M_{\rm bound,sat}/M_{\rm 200,host}) < -3$), while the bottom panel displays the remaining satellites. Coloured solid lines indicate the median trends of different bins of mass ratios (as in Fig. \ref{SMSR_sats_COLIBRE}), and black dotted lines mark the median, 16th, and 84th percentiles for central galaxies. Pink and dark green squares in the top panel denote the median values of the ``dwarf'' and ``bright'' subsamples, respectively. Bright satellites exhibit higher metallicities and are relatively more baryon-dominated than dwarf satellites, which are typically more metal-poor.}
    \label{FeH_Mstar_fDM_Sat_z0}
\end{figure}

\begin{figure*}
    \centering
    \includegraphics[width=1.9\columnwidth]{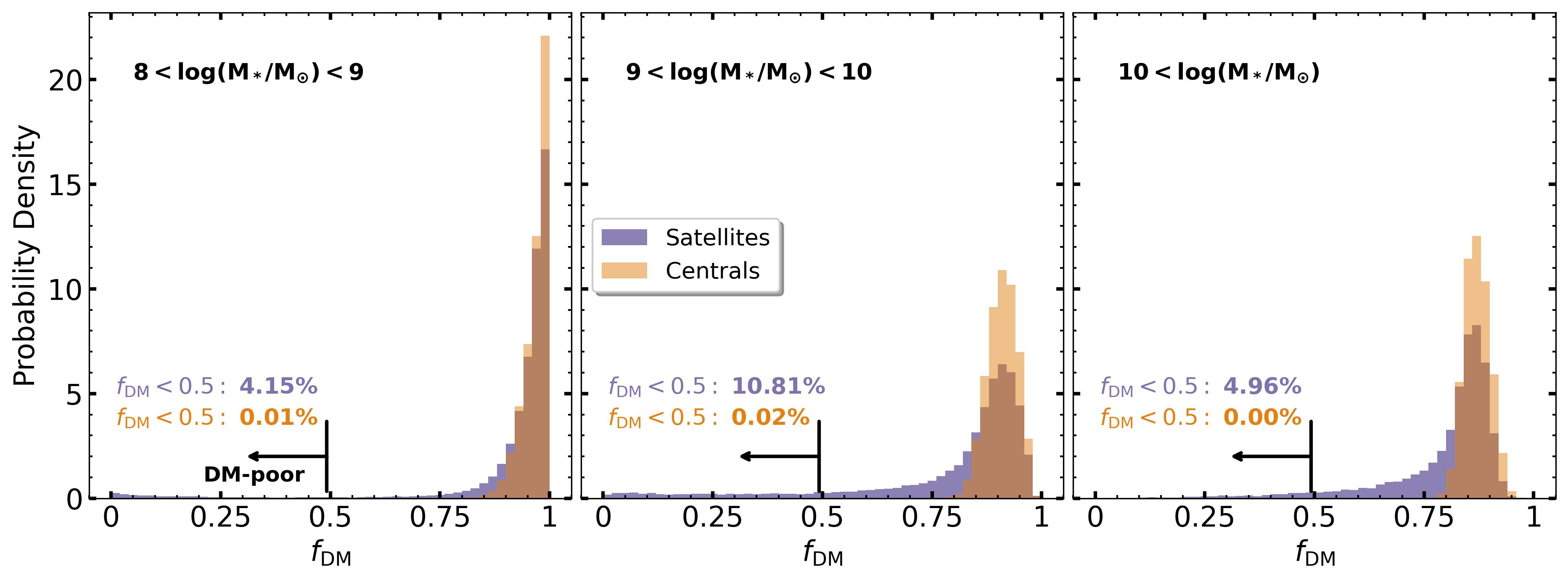}
    
    \caption{Distribution of DM mass fractions ($f_{\rm DM}$) of satellites (purple) and centrals (orange) in three different bins of $z=0$ stellar mass. Values are measured using all bound particles within 50 kpc of each galaxy. From left to right: $8 < \log (M_{*}/{\rm M}_{\odot}) < 9$, $9 < \log (M_{*}/{\rm M}_{\odot}) < 10$, and $10 < \log (M_{*}/{\rm M}_{\odot})$. The percentage of galaxies that can be considered ``DM-poor'' ($f_{\rm DM} < 0.5$, black arrow) in each subsample is indicated in the legend.}
    \label{fDM_distrib}
\end{figure*}

\subsection{Satellite evolution in mass and size}
\label{subsec:SMSR_TemporalEvolution}

We study here the mass and size evolution of satellites since their infall. Motivated by the preceding discussion, we focus here on satellites with very low mass ratios, i.e. $\log(M_{\rm bound,sat}/M_{\rm 200,host}) < -3$, which were accreted at a similar cosmic time ($6 < t_{\rm infall}/{\rm Gyr} < 8$), and survive to $z=0$.

Fig. \ref{SMSR_tracks_subsamples} shows how the stellar mass and half-mass radius of this population change after infall, and compares them with the corresponding evolution of centrals of comparable stellar mass. The left panel shows the median evolution (solid lines) of central galaxies, grouped in two logarithmic bins of $z=0$ log stellar mass ($[8, 8.8]$ and $[9.5,10]$, in solar units), between $t = 6$ Gyr (circles) and $z=0$ (squares). Galaxies in these mass bins are seen to grow both in mass and size, and track the median evolution of the centrals' SMSR.

Satellites in the same $z=0$ mass bins, on the other hand, evolve differently, and deviate from the trend that holds for centrals at $z=0$. These galaxies were selected to sample the median satellite SMSR at $z=0$ (purple solid line), with error bars indicating the region spanned in the mass-size plane by each subsample. By tracking these surviving systems backward in time, we find that their pre-infall sizes and stellar masses were entirely representative of the average central galaxy population at that epoch. This demonstrates that their different present-day sizes relative to present-day centrals are driven by post-infall physical evolution rather than survivor selection bias.

Dwarf satellites grow mainly in size (by more than a factor of 2 in the median), but little in stellar mass (shown in pink in the middle panel of Fig. \ref{SMSR_tracks_subsamples}). On the other hand, bright satellites (dark green) evolve mainly in stellar mass, keeping, on average, their original size.

Lastly, the right panel of Fig. \ref{SMSR_tracks_subsamples} traces back in time the most extreme satellites at $z=0$ in the range $8.5 < \log (M_{*}/{\rm M}_{\odot}) < 9.5$ (in terms of size at fixed stellar mass, top and bottom 5th percentile). Interestingly, the most compact satellites in this mass range are formed by systems which were once larger and more massive than at present. The largest satellites, on the other hand, expand more in size than in mass, following a more extreme evolution than other dwarf satellites.

These results suggest a simple interpretation for the distinct SMSR of satellites shown in Fig. \ref{SMSR_sats_COLIBRE}. Dwarf satellites ($8 < \log (M_{*}/{\rm M}_{\odot}) < 9$) grow preferentially in size rather than in mass, whereas the opposite appears true for bright satellites ($9 < \log (M_{*}/{\rm M}_{\odot}) < 10$), leading to the ``wiggle'' seen in the satellite SMSR at $z=0$.

Given the DM super-sampling implemented in COLIBRE, one would expect a priori that the size growth of low-mass satellites does not stem from numerical artifacts, such as spurious energy transfer to stellar particles \citep[see][]{Ludlow2023}. We explore below the reason for these distinct evolutionary trends of dwarf and bright satellites.

\subsubsection{Dwarf satellites}
\label{EvDwarfSat}

The evolutionary trends shown in Fig. \ref{f_Survivors_t_infall} and Fig. \ref{SMSR_tracks_subsamples} suggest a possible interpretation for the evolution of dwarf satellites in the stellar mass-size plane. These figures show that, after infall, many of these satellites are fully disrupted; those that survive to $z=0$ retain their stellar mass at infall but systematically increase in size, often substantially. 

This is consistent with a scenario in which dwarf satellites lose their gas content soon after infall due to ram pressure exerted by a hot gaseous atmosphere around of the host (thus curtailing their ability to further form stars), and whose sizes are gradually increased by tides until they fully disrupt. Satellites with steep inner density profiles shrink rather than expand as a result of tides because the outskirts of the system are lost first \citep[see e.g.][]{Penarrubia2008, Penarrubia2010, Errani2024}. However, this is not the case when a system has a constant density core in its inner density profile and full satellite disruption can occur \citep[see e.g.][]{Penarrubia2012, Carleton2019, Errani23}. In this case, the density of a satellite is continually eroded, expanding as it approaches full disruption. A more detailed analysis of the tidal evolutionary tracks of COLIBRE satellites is presented in Appendix~\ref{AppTET}.

This interpretation is supported by the pre-infall (at $t\approx 6 ~ {\rm Gyr}$) mass profile of dwarf satellites, shown in the top panel of Fig.~\ref{stack_hydro_vs_DMO}. Here, the red (stars) and green (gas) profiles indicate that the baryonic component contributes much less than the dark matter (dotted orange curve) to the total mass profile, except at scales below the gravitational softening length, where the gas component reaches comparable densities.

These dwarf satellites lose their gas soon after infall, as shown in the top panel of Fig.~\ref{evo_temp_subsamples_props}, which tracks the gas and stellar mass within $r < 3$ kpc of these systems after infall. Over $\sim$1 Gyr after infall, gas is removed and star formation is terminated. This rapid gas loss causes the central gravitational potential well to shallow, prompting an immediate expansion of the stellar distribution due to dynamical relaxation. As shown in the top panel of Fig. \ref{evo_temp_subsamples_props}, this initial impulsive relaxation accounts for roughly half of the total size growth observed after infall (a $\approx$30\% increase in $r_{\rm h,*}$). However, gas loss alone cannot explain the full galaxy expansion, as the stellar size continues to expand gradually over several subsequent Gyr, tracking the secular tidal stripping of the remaining enclosed dark matter mass.

Once the gas component has been lost, the tidal fate of these satellites is completely determined by their DM profiles. At the time of infall, $\sim$75\% of the total mass in the inner 3 pkpc is DM. As seen in Fig.~\ref{stack_hydro_vs_DMO}, the dark matter central profile exhibits a rather large ($\sim$2 kpc) nearly constant density core. This core size is a factor $\sim$3 larger than the softening length of the simulation (vertical grey dotted line), and is substantially larger than the numerical cores of subhaloes with similar mass from the DMO simulation, as shown by the black curve in Fig.~\ref{stack_hydro_vs_DMO}. The large reduction in the central dark matter mass experienced by the halo in the COLIBRE hydro simulation is most likely a result of energy transfer between dark matter and baryons during the assembly of the central galaxy \citep[see e.g.][]{Navarro1996b,Governato2012,Penarrubia2012,BL2019,Lazar2020}.

The large cores shown in Fig.~\ref{stack_hydro_vs_DMO} exacerbate the tidal disruption of the satellite, which explains why tides gradually increase the size of the stellar component of dwarf satellites (red solid line in Fig. \ref{evo_temp_subsamples_props}) until they are eventually fully disrupted. As shown in Fig. \ref{f_Survivors_t_infall}, this is not a subtle effect; more than $70\%$ of dwarf satellites accreted at $t \sim 6$ Gyr (i.e., roughly $8$ Gyrs ago, at $z=1$) do not survive to $z=0$. 

\subsubsection{Bright satellites}
\label{Evbrightat}

A similar exercise also explains why bright satellites evolve differently from dwarfs. Their median mass profiles, again for systems that are accreted onto their main hosts at $t\sim 6$ Gyr, are shown at the time of accretion in the bottom panel of Fig.~\ref{stack_hydro_vs_DMO}.
The differences with the median dwarf satellite profile (top panel) are striking. Although their DM profiles are similar (not unexpected, since they differ in halo mass by just over a factor of two), their baryonic profiles are quite different, and dominate over the DM in the inner regions. The baryonic domination over the DM in the inner regions is further enhanced by the 
core in the DM inner profile, which reduces the inner DM content relative to the DMO case. At the time of infall, only $\approx 50$\% of the total mass in the innermost 3 pkpc is DM, as shown in Fig. \ref{evo_temp_subsamples_props}.

These baryon-dominated satellites evolve quite differently. Both the gas and stellar components of bright satellites are much denser than those of dwarf satellites, which makes them much more resilient to tides within the same potential. This is shown in the bottom panel of Fig.~\ref{evo_temp_subsamples_props}, where we see that tidal effects after infall cannot efficiently remove the most bound baryons. Instead, these satellites retain their central gas reservoir, where ram-pressure compression can further promote compact star formation in the inner regions \citep[see e.g.][]{Du2019,Steyrleithner2020}. This is reflected in the steady increase of enclosed stellar mass within $r < 3$ kpc post-infall, converting inner gas into stars while keeping the overall stellar size roughly constant. Bright satellites grow in stellar mass by a factor of $\sim$2-3 on average while maintaining a roughly constant size, resulting from the balance between concentrated central star formation and the tidal heating and expansion of their older stellar components (see also Appendix~\ref{AppTET} for details). 

Although bright satellites are structurally resilient, roughly 59\% of those accreted at $t \sim 6$ Gyr do not survive to the present day as distinct subhaloes (Fig. \ref{f_Survivors_t_infall}). This non-survival is primarily driven by efficient dynamical friction sinking massive satellites toward the host centre, leading to mergers, as well as the eventual disruption of systems that lose their extended dark matter envelopes to tides. Because tidal stellar mass loss of a subhalo only becomes effective after it loses more than $\approx$95\% of its originally bound mass \citep{He2026}, survivor selection naturally leads the compact, baryon-dominated nature of the surviving population at $z=0$.

\subsection{Mass-metallicity relation of satellites and centrals}
\label{subsec:SatPropsObs}

In the scenario presented above, dwarf satellites at $z=0$ are typically larger than centrals because they lose their gravitationally-dominant gas quickly, stop forming stars soon after infall, and expand on their way to being tidally disrupted. This differential evolution of satellites relative to centrals should also result in differences in the stellar mass-metallicity relation (SMMR), which has been extensively studied for centrals \citep[see e.g.][]{DekelSilk1986,Matteucci1994,Pasquali2012,Kirby2013,Weisz2015,Sharda2026}. 

The lack of star formation after infall, in particular, implies that dwarf satellites should track the mass-metallicity relation of centrals at their time of accretion. Since centrals generally keep forming stars and enriching further, we expect them to have higher metallicities today than quenched dwarf satellites at fixed $M_*$.

Bright satellites, on the other hand, retain their inner gas and continue forming stars after infall, but without further gas accretion to replenish their gas reservoirs \citep[see e.g.][]{Wright2022}. They should then self-enrich to metallicities higher than those of centrals of comparable mass, whose metallicity enrichment would have been somewhat diluted by the continuous accretion of external gas \citep[e.g.][]{Cole2000,Bahe2017,Du2019,Gallazzi2021}.

This would result in a ``wiggle'' in the stellar mass-metallicity relation (SMMR) of satellites relative to centrals, akin to the one seen for the SMSR relation. We verify this in Fig. \ref{FeH_Mstar_fDM_Sat_z0}, where we show the mass-weighted iron-to-hydrogen stellar abundances (in solar units, ${\rm [Fe/H]}$), calculated from all bound stellar particles within 50 kpc, as a function of stellar mass for all $z=0$ satellites. The dotted black line traces the SMMR for centrals at $z=0$, with thinner lines indicating the 16th and 84th percentiles. Solid coloured lines correspond to satellites of different $M_{\rm bound,sat}/M_{\rm 200,host}$ (as in Fig. \ref{SMSR_sats_COLIBRE}), and exhibit a ``wiggle'' relative to the centrals' SMMR, as anticipated above. The effect is larger for lower mass ratios.

Small circles in Fig. \ref{FeH_Mstar_fDM_Sat_z0} correspond to individual satellites; most track the median trend shown by the solid coloured lines. However, for systems with the lowest mass ratios (top panel), a significant population scatters above the main relation, occasionally reaching very high metallicities for their stellar mass. 

Where do such systems come from? A clue to their origin is provided by their dark matter mass fraction, $f_{\rm DM}=M_{\rm DM}/M_{\rm tot}$ (measured using all bound particles within a 50 pkpc aperture), which is used to colour code individual satellites. The colour bar on the right-hand side of Fig. \ref{FeH_Mstar_fDM_Sat_z0} indicates that these outliers are systems which have lost most of their dark matter component; their $f_{\rm DM}$ is less than 25\% (and in some cases zero), whereas most systems on the main SMMR have $f_{\rm DM}>0.75$.

These ``dark matter-poor'', high-metallicity satellites may be traced back to systems which, at infall, had highly concentrated, gravitationally dominant baryonic components, extreme examples of the relative mass distribution presented in the bottom panel of Fig.~\ref{stack_hydro_vs_DMO}. In these cases, the core allows the dark matter component to be fully stripped, leaving behind a self-bound, gas-rich, highly concentrated baryon ``nugget'' which continues to form stars and self-enrich to high metallicity until the gas component is exhausted.

This picture is similar to the recent findings of \citet{deAlmeida2026}, who studied dwarf satellite subsamples in TNG50-1 and report that DM-poor satellites are typically smaller and more metal-rich due to a combination of stripping of the outer stellar component and starbursts in the innermost regions driven by ram pressure after accretion.

In COLIBRE, these metal-rich outliers ($\approx$15\% of all satellites with $\log(M_{\rm bound,sat}/M_{\rm 200,host}) < -3$) are satellites that were once much more massive, but where tidal stripping has led to the loss of the dark matter component and of a large fraction of their stars \citep[see][]{He2026}, shifting them to a region of the SMMR plane of much higher metallicity than expected from their current stellar mass. In contrast, fewer than 1\% of systems with higher mass ratios ($\log(M_{\rm bound,sat}/M_{\rm 200,host}) > -3$, bottom panel) exhibit $f_{\rm DM} < 0.5$, making DM-poor satellites statistically negligible at higher mass ratios.

The existence of these outliers in the satellite SMMR is an intriguing prediction of the COLIBRE simulation which may, in principle, be tested observationally. For that purpose, it is important to establish the fraction of satellites of given $M_*$ that are ``DM poor''. This is illustrated in Fig.~\ref{fDM_distrib}, which shows the distribution of DM mass fractions for both centrals and satellites across three bins of $M_*$. Dark matter-poor galaxies are not particularly rare in COLIBRE, making up to $\sim$11\% of all satellites with $9 < \log(M_*/{\rm M}_\odot) < 10$ at $z=0$ \citep[see also][]{He2026}. Finding and characterizing this population in observations would provide strong support for the applicability of COLIBRE results to actual satellites. A more detailed investigation into the origins of these systems and detailed comparison with existing observational data \citep[e.g.][]{vanDokkum2018,vanDokkum2019,vanDokkum2022,Mercado2025} will be presented in future work.

\section{Summary and conclusions} \label{SecConc}

We have used the L200m6 simulation from the COLIBRE suite of hydrodynamic cosmological simulations to investigate the distinct evolutionary paths induced on satellite galaxies by their environment and, in particular, the origin of differences in the Stellar Mass-Size Relation (SMSR) of satellites and centrals.

We find that, at $z=0$, the satellite SMSR in COLIBRE ``wiggles'' about that of centrals (Fig. \ref{SMSR_sats_vs_cent}): dwarf satellites ($8 < \log (M_{*}/{\rm M}_{\odot}) < 9$) are typically more extended than centrals of the same stellar mass, a trend that reverses for bright satellites ($9 < \log (M_{*}/{\rm M}_{\odot}) < 10.5$). 
This difference is particularly pronounced for the least massive satellites, as measured by their satellite-to-host total mass ratios (Fig. \ref{SMSR_sats_COLIBRE}).

The distinct SMSR of satellites can be explained by the radial distribution of DM and baryons at infall. Dwarf satellites are gas-rich, DM dominated systems which lose most of their gas soon after infall (Fig. \ref{evo_temp_subsamples_props}). Their dark matter inner profile exhibits a relatively large core (Fig. \ref{stack_hydro_vs_DMO}), most likely induced by the infall and sudden removal of baryons from the inner regions during the early formation of the galaxy \citep[see e.g.,][]{Governato2010, Governato2012,DiCintio2014,Christensen2016,BL2019}, but given the limited resolution of the simulation, numerical relaxation cannot be fully excluded \citep{Ludlow2019}. These cores make dwarf systems particularly susceptible to full tidal disruption. Following the initial expansion triggered by rapid gas loss, continuing tidal stripping gradually increases their stellar sizes with little change in stellar mass, explaining why dwarf satellites are larger than centrals of comparable stellar mass at $z=0$.

Bright satellites, on the other hand, are much denser, and stars dominate the potential well at infall, being thus much more resilient to tides. As a result, they can retain their gas and stars and keep forming stars after infall. These satellites cannot accrete further gas, so their star formation is restricted to the inner regions. As a result, their stellar masses grow while their sizes remain nearly constant (Fig. \ref{SMSR_tracks_subsamples}), explaining why bright satellites are smaller than centrals of comparable mass at $z=0$.

The same scenario explains why the stellar mass-metallicity relation (SMMR) of COLIBRE satellites is found to ``wiggle'' about that of centrals. Dwarf satellites are more metal-poor, whereas bright satellites are more metal-rich than centrals of comparable mass, although the overall effect is small (Fig. \ref{FeH_Mstar_fDM_Sat_z0}).

We also find that a significant fraction of bright satellites in COLIBRE end up being  ``dark-matter poor'' or even ``dark-matter free'' (Fig. \ref{fDM_distrib}); i.e., they lose much or all of their dark matter component to tides, but leave behind a tight, self-bound collection of stars. This DM-poor satellite population is particularly evident in the mass-metallicity relation, producing systems that are systematically more metal-rich than expected for their stellar mass.

Finally, the presence of dark matter cores in dwarfs makes them particularly vulnerable to complete tidal disruption: only $34\%$ of dwarf satellites accreted at $6 < t_{\rm infall}/{\rm Gyr} < 8$ (i.e., $z \approx 1$) survive to $z=0$.  For bright satellites, the non-survival fraction (59\%, Fig. \ref{f_Survivors_t_infall}) reflects the combined impact of tidal mass loss and orbital decay driven by dynamical friction leading to mergers with the host.

These results provide a theoretical foundation for interpreting satellite diversity and offer testable predictions for deep-imaging surveys. Specifically, our framework predicts that low-mass, extended galaxies are systems undergoing tidal disruption, whereas compact galaxies are the more metal-rich, DM-deficient remnants of previously more massive galaxies. Ultimately, this work suggests that the interplay between internal baryonic and DM distributions within strong tidal fields is a primary driver of satellite structural diversity in a cosmological context.

\begin{acknowledgements}
    This project has received funding from the European Union’s HORIZON-MSCA-2021-SE-01 Research and Innovation programme under the Marie Sklodowska-Curie grant agreement number 101086388 - Project acronym: LACEGAL. This work was partially supported by the Consejo de Investigaciones Científicas y Técnicas de la República Argentina (CONICET) and the Secretaría de Ciencia y Técnica de la Universidad Nacional de Córdoba (SeCyT). We acknowledge the use of the SwiftSimIO open source tools \citep{Borrow2020}. This work used the DiRAC@Durham facility managed by the Institute for Computational Cosmology on behalf of the STFC DiRAC HPC Facility (\url{www.dirac.ac.uk}). The equipment was funded by BEIS capital funding via STFC capital grants ST/K00042X/1, ST/P002293/1, ST/R002371/1 and ST/S002502/1, Durham University and STFC operations grant ST/R000832/1. DiRAC is part of the National e-Infrastructure. This project has received funding from the Netherlands Organization for Scientific Research (NWO) through research programme Athena 184.034.002. JFN acknowledges the hospitality of Durham University during the completion of this work. ABL acknowledges support by the Italian Ministry for Universities (MUR) program “Dipartimenti di Eccellenza 2023-2027” within the Centro Bicocca di Cosmologia Quantitativa (BiCoQ), and support by UNIMIB’s Fondo Di Ateneo Quota Competitiva (project 2024-ATEQC-0050). SP acknowledges support by the Austrian Science Fund (FWF) through grant-DOI: 10.55776/V982. 
\end{acknowledgements}

\bibliographystyle{aa}
\bibliography{example}

@ARTICLE{Errani23,
       author = {{Errani}, Rapha{\"e}l and {Navarro}, Julio F. and {Pe{\~n}arrubia}, Jorge and {Famaey}, Benoit and {Ibata}, Rodrigo},
        title = "{Dark matter halo cores and the tidal survival of Milky Way satellites}",
      journal = {\mnras},
         year = 2023,
        month = feb,
       volume = {519},
       number = {1},
        pages = {384-396},
          doi = {10.1093/mnras/stac3499},
archivePrefix = {arXiv},
       eprint = {2210.01131},
 primaryClass = {astro-ph.GA},
       adsurl = {https://ui.adsabs.harvard.edu/abs/2023MNRAS.519..384E}
}

@ARTICLE{Schaye2025,
       author = {{Schaye}, Joop and {Chaikin}, Evgenii and {Schaller}, Matthieu and {Ploeckinger}, Sylvia and {Hu{\v{s}}ko}, Filip and {McGibbon}, Robert J. and {Trayford}, James W. and {Ben{\'\i}tez-Llambay}, Alejandro and {Correa}, Camila and {Frenk}, Carlos S. and {Richings}, Alexander J. and {Forouhar Moreno}, Victor J. and {Bah{\'e}}, Yannick M. and {Borrow}, Josh and {Durrant}, Anna and {Gebek}, Andrea and {Helly}, John C. and {Jenkins}, Adrian and {Lacey}, Cedric G. and {Ludlow}, Aaron and {Nobels}, Folkert S.~J.},
        title = "{The COLIBRE project: cosmological hydrodynamical simulations of galaxy formation and evolution}",
      journal = {\mnras},
         year = 2026,
        month = may,
       volume = {548},
       number = {1},
          eid = {stag375},
        pages = {stag375},
          doi = {10.1093/mnras/stag375},
archivePrefix = {arXiv},
       eprint = {2508.21126},
 primaryClass = {astro-ph.GA},
       adsurl = {https://ui.adsabs.harvard.edu/abs/2026MNRAS.548ag375S}
}

@ARTICLE{Ludlow2026,
       author = {{Ludlow}, Aaron D. and {Proctor}, Katy L. and {Schaye}, Joop and {Hu{\v{s}}ko}, Filip and {Forouhar Moreno}, Victor J. and {Obreschkow}, Danail and {Chaikin}, Evgenii and {Schaller}, Matthieu and {Ploeckinger}, Sylvia and {Ben{\'\i}tez-Llambay}, Alejandro and {Oman}, Kyle A. and {McGibbon}, Robert J. and {Trayford}, James W. and {Frenk}, Carlos S. and {Richings}, Alexander J.},
        title = "{The evolution of the sizes and angular momentum content of galaxies in the COLIBRE simulations}",
      journal = {arXiv e-prints},
         year = 2026,
        month = mar,
          eid = {arXiv:2603.26200},
        pages = {arXiv:2603.26200},
          doi = {10.48550/arXiv.2603.26200},
archivePrefix = {arXiv},
       eprint = {2603.26200},
 primaryClass = {astro-ph.GA},
       adsurl = {https://ui.adsabs.harvard.edu/abs/2026arXiv260326200L}
}

@ARTICLE{Ludlow2019,
       author = {{Ludlow}, Aaron D. and {Schaye}, Joop and {Bower}, Richard},
        title = "{Numerical convergence of simulations of galaxy formation: the abundance and internal structure of cold dark matter haloes}",
      journal = {\mnras},
         year = 2019,
        month = sep,
       volume = {488},
       number = {3},
        pages = {3663-3684},
          doi = {10.1093/mnras/stz1821},
archivePrefix = {arXiv},
       eprint = {1812.05777},
 primaryClass = {astro-ph.CO},
       adsurl = {https://ui.adsabs.harvard.edu/abs/2019MNRAS.488.3663L}
}

@ARTICLE{Penarrubia2012,
       author = {{Pe{\~n}arrubia}, Jorge and {Pontzen}, Andrew and {Walker}, Matthew G. and {Koposov}, Sergey E.},
        title = "{The Coupling between the Core/Cusp and Missing Satellite Problems}",
      journal = {\apjl},
         year = 2012,
        month = nov,
       volume = {759},
       number = {2},
          eid = {L42},
        pages = {L42},
          doi = {10.1088/2041-8205/759/2/L42},
archivePrefix = {arXiv},
       eprint = {1207.2772},
 primaryClass = {astro-ph.GA},
       adsurl = {https://ui.adsabs.harvard.edu/abs/2012ApJ...759L..42P}
}

@ARTICLE{Penarrubia2010,
       author = {{Pe{\~n}arrubia}, Jorge and {Benson}, Andrew J. and {Walker}, Matthew G. and {Gilmore}, Gerard and {McConnachie}, Alan W. and {Mayer}, Lucio},
        title = "{The impact of dark matter cusps and cores on the satellite galaxy population around spiral galaxies}",
      journal = {\mnras},
         year = 2010,
        month = aug,
       volume = {406},
       number = {2},
        pages = {1290-1305},
          doi = {10.1111/j.1365-2966.2010.16762.x},
archivePrefix = {arXiv},
       eprint = {1002.3376},
 primaryClass = {astro-ph.GA},
       adsurl = {https://ui.adsabs.harvard.edu/abs/2010MNRAS.406.1290P}
}

@ARTICLE{Governato2012,
       author = {{Governato}, F. and {Zolotov}, A. and {Pontzen}, A. and {Christensen}, C. and {Oh}, S.~H. and {Brooks}, A.~M. and {Quinn}, T. and {Shen}, S. and {Wadsley}, J.},
        title = "{Cuspy no more: how outflows affect the central dark matter and baryon distribution in {\ensuremath{\Lambda}} cold dark matter galaxies}",
      journal = {\mnras},
         year = 2012,
        month = may,
       volume = {422},
       number = {2},
        pages = {1231-1240},
          doi = {10.1111/j.1365-2966.2012.20696.x},
archivePrefix = {arXiv},
       eprint = {1202.0554},
 primaryClass = {astro-ph.CO},
       adsurl = {https://ui.adsabs.harvard.edu/abs/2012MNRAS.422.1231G}
}

@ARTICLE{Bahe2017,
       author = {{Bah{\'e}}, Yannick M. and {Schaye}, Joop and {Crain}, Robert A. and {McCarthy}, Ian G. and {Bower}, Richard G. and {Theuns}, Tom and {McGee}, Sean L. and {Trayford}, James W.},
        title = "{The origin of the enhanced metallicity of satellite galaxies}",
      journal = {\mnras},
         year = 2017,
        month = jan,
       volume = {464},
       number = {1},
        pages = {508-529},
          doi = {10.1093/mnras/stw2329},
archivePrefix = {arXiv},
       eprint = {1609.03379},
 primaryClass = {astro-ph.GA},
       adsurl = {https://ui.adsabs.harvard.edu/abs/2017MNRAS.464..508B}
}

@ARTICLE{He2026,
       author = {{He}, Feihong and {Han}, Jiaxin and {Schaye}, Joop and {Wang}, Wenting and {Li}, Zhaozhou and {Ploeckinger}, Sylvia and {Chaikin}, Evgenii and {McGibbon}, Robert J. and {Hu{\v{s}}ko}, Filip and {Schaller}, Matthieu and {Ben{\'\i}tez-Llambay}, Alejandro and {Richings}, Alexander J. and {Trayford}, James W. and {Frenk}, Carlos S. and {Jiang}, Fangzhou},
        title = "{The tidal evolution of satellite galaxies in cosmological simulations: insights from COLIBRE}",
      journal = {\mnras},
         year = 2026,
        month = sep,
       volume = {551},
       number = {1},
          eid = {stag1395},
        pages = {stag1395},
          doi = {10.1093/mnras/stag1395},
archivePrefix = {arXiv},
       eprint = {2604.03105},
 primaryClass = {astro-ph.GA},
       adsurl = {https://ui.adsabs.harvard.edu/abs/2026MNRAS.551g1395H}
}

@ARTICLE{Forouhar2026,
       author = {{Forouhar Moreno}, Victor J. and {Schaye}, Joop and {Schaller}, Matthieu and {Ludlow}, Aaron and {McGibbon}, Robert J. and {Ben{\'\i}tez-Llambay}, Alejandro and {Chaikin}, Evgenii and {Frenk}, Carlos S. and {Hu{\v{s}}ko}, Filip and {Ploeckinger}, Sylvia and {Richings}, Alexander J. and {Trayford}, James W.},
        title = "{The morphologies of present-day galaxies in the COLIBRE simulations}",
      journal = {arXiv e-prints},
         year = 2026,
        month = apr,
          eid = {arXiv:2604.03503},
        pages = {arXiv:2604.03503},
          doi = {10.48550/arXiv.2604.03503},
archivePrefix = {arXiv},
       eprint = {2604.03503},
 primaryClass = {astro-ph.GA},
       adsurl = {https://ui.adsabs.harvard.edu/abs/2026arXiv260403503F}
}

@ARTICLE{Chaikin2025,
       author = {{Chaikin}, Evgenii and {Schaye}, Joop and {Schaller}, Matthieu and {Ploeckinger}, Sylvia and {Ben{\'\i}tez-Llambay}, Alejandro and {Frenk}, Carlos S. and {Hu{\v{s}}ko}, Filip and {McGibbon}, Robert J. and {Richings}, Alexander J. and {Trayford}, James W.},
        title = "{The evolution of the galaxy stellar mass function and star formation rates in the COLIBRE simulations from redshift 17 to 0}",
      journal = {\mnras},
         year = 2026,
        month = jun,
       volume = {548},
       number = {4},
          eid = {stag740},
        pages = {stag740},
          doi = {10.1093/mnras/stag740},
archivePrefix = {arXiv},
       eprint = {2509.07960},
 primaryClass = {astro-ph.GA},
       adsurl = {https://ui.adsabs.harvard.edu/abs/2026MNRAS.548ag740C}
}

@ARTICLE{Larson1980,
       author = {{Larson}, R.~B. and {Tinsley}, B.~M. and {Caldwell}, C.~N.},
        title = "{The evolution of disk galaxies and the origin of S0 galaxies}",
      journal = {\apj},
         year = 1980,
        month = may,
       volume = {237},
        pages = {692-707},
          doi = {10.1086/157917},
       adsurl = {https://ui.adsabs.harvard.edu/abs/1980ApJ...237..692L}
}

@ARTICLE{Sawala2016dwarfs,
       author = {{Sawala}, Till and {Frenk}, Carlos S. and {Fattahi}, Azadeh and {Navarro}, Julio F. and {Theuns}, Tom and {Bower}, Richard G. and {Crain}, Robert A. and {Furlong}, Michelle and {Jenkins}, Adrian and {Schaller}, Matthieu and {Schaye}, Joop},
        title = "{The chosen few: the low-mass haloes that host faint galaxies}",
      journal = {\mnras},
         year = 2016,
        month = feb,
       volume = {456},
       number = {1},
        pages = {85-97},
          doi = {10.1093/mnras/stv2597},
archivePrefix = {arXiv},
       eprint = {1406.6362},
 primaryClass = {astro-ph.CO},
       adsurl = {https://ui.adsabs.harvard.edu/abs/2016MNRAS.456...85S}
}

@ARTICLE{PressDavis1982,
       author = {{Press}, W.~H. and {Davis}, M.},
        title = "{How to identify and weigh virialized clusters of galaxies in a complete redshift catalog}",
      journal = {\apj},
         year = 1982,
        month = aug,
       volume = {259},
        pages = {449-473},
          doi = {10.1086/160183},
       adsurl = {https://ui.adsabs.harvard.edu/abs/1982ApJ...259..449P}
}

@ARTICLE{Font2022,
       author = {{Font}, Andreea S. and {McCarthy}, Ian G. and {Belokurov}, Vasily and {Brown}, Shaun T. and {Stafford}, Sam G.},
        title = "{Quenching of satellite galaxies of Milky Way analogues: reconciling theory and observations}",
      journal = {\mnras},
         year = 2022,
        month = mar,
       volume = {511},
       number = {1},
        pages = {1544-1556},
          doi = {10.1093/mnras/stac183},
archivePrefix = {arXiv},
       eprint = {2109.06215},
 primaryClass = {astro-ph.GA},
       adsurl = {https://ui.adsabs.harvard.edu/abs/2022MNRAS.511.1544F}
}

@ARTICLE{Font2021,
       author = {{Font}, Andreea S. and {McCarthy}, Ian G. and {Belokurov}, Vasily},
        title = "{Can cosmological simulations capture the diverse satellite populations of observed Milky Way analogues?}",
      journal = {\mnras},
         year = 2021,
        month = jul,
       volume = {505},
       number = {1},
        pages = {783-801},
          doi = {10.1093/mnras/stab1332},
archivePrefix = {arXiv},
       eprint = {2011.12974},
 primaryClass = {astro-ph.GA},
       adsurl = {https://ui.adsabs.harvard.edu/abs/2021MNRAS.505..783F}
}

@ARTICLE{SteinmetzWhite1997,
       author = {{Steinmetz}, Matthias and {White}, Simon D.~M.},
        title = "{Two-body heating in numerical galaxy formation experiments}",
      journal = {\mnras},
         year = 1997,
        month = jul,
       volume = {288},
       number = {3},
        pages = {545-550},
          doi = {10.1093/mnras/288.3.545},
archivePrefix = {arXiv},
       eprint = {astro-ph/9609021},
 primaryClass = {astro-ph},
       adsurl = {https://ui.adsabs.harvard.edu/abs/1997MNRAS.288..545S}
}

@ARTICLE{Steyrleithner2020,
       author = {{Steyrleithner}, P. and {Hensler}, G. and {Boselli}, A.},
        title = "{The effect of ram-pressure stripping on dwarf galaxies}",
      journal = {\mnras},
         year = 2020,
        month = may,
       volume = {494},
       number = {1},
        pages = {1114-1127},
          doi = {10.1093/mnras/staa775},
archivePrefix = {arXiv},
       eprint = {2003.09591},
 primaryClass = {astro-ph.GA},
       adsurl = {https://ui.adsabs.harvard.edu/abs/2020MNRAS.494.1114S}
}

@ARTICLE{Du2019,
       author = {{Du}, Min and {Debattista}, Victor P. and {Ho}, Luis C. and {C{\^o}t{\'e}}, Patrick and {Spengler}, Chelsea and {Erwin}, Peter and {Wadsley}, James W. and {Norris}, Mark A. and {Earp}, Samuel W.~F. and {Quinn}, Thomas R. and {Fiteni}, Karl and {Caruana}, Joseph},
        title = "{The Formation of Compact Elliptical Galaxies in the Vicinity of a Massive Galaxy: The Role of Ram-pressure Confinement}",
      journal = {\apj},
         year = 2019,
        month = apr,
       volume = {875},
       number = {1},
          eid = {58},
        pages = {58},
          doi = {10.3847/1538-4357/ab0e0c},
archivePrefix = {arXiv},
       eprint = {1811.06778},
 primaryClass = {astro-ph.GA},
       adsurl = {https://ui.adsabs.harvard.edu/abs/2019ApJ...875...58D}
}

@ARTICLE{McCarthy2008,
       author = {{McCarthy}, I.~G. and {Frenk}, C.~S. and {Font}, A.~S. and {Lacey}, C.~G. and {Bower}, R.~G. and {Mitchell}, N.~L. and {Balogh}, M.~L. and {Theuns}, T.},
        title = "{Ram pressure stripping the hot gaseous haloes of galaxies in groups and clusters}",
      journal = {\mnras},
         year = 2008,
        month = jan,
       volume = {383},
       number = {2},
        pages = {593-605},
          doi = {10.1111/j.1365-2966.2007.12577.x},
archivePrefix = {arXiv},
       eprint = {0710.0964},
 primaryClass = {astro-ph},
       adsurl = {https://ui.adsabs.harvard.edu/abs/2008MNRAS.383..593M}
}

@ARTICLE{Cole2000,
       author = {{Cole}, Shaun and {Lacey}, Cedric G. and {Baugh}, Carlton M. and {Frenk}, Carlos S.},
        title = "{Hierarchical galaxy formation}",
      journal = {\mnras},
         year = 2000,
        month = nov,
       volume = {319},
       number = {1},
        pages = {168-204},
          doi = {10.1046/j.1365-8711.2000.03879.x},
archivePrefix = {arXiv},
       eprint = {astro-ph/0007281},
 primaryClass = {astro-ph},
       adsurl = {https://ui.adsabs.harvard.edu/abs/2000MNRAS.319..168C}
}

@ARTICLE{Correa2026,
       author = {{Correa}, Camila A. and {Schaye}, Joop and {Schaller}, Matthieu and {Trayford}, James W. and {Chaikin}, Evgenii and {Ben{\'\i}tez-Llambay}, Alejandro and {Frenk}, Carlos S. and {Ploeckinger}, Sylvia and {Richings}, Alexander J.},
        title = "{A subgrid model for chemical enrichment in cosmological simulations}",
      journal = {\mnras},
         year = 2026,
        month = may,
       volume = {548},
       number = {3},
          eid = {stag645},
        pages = {stag645},
          doi = {10.1093/mnras/stag645},
archivePrefix = {arXiv},
       eprint = {2604.00980},
 primaryClass = {astro-ph.GA},
       adsurl = {https://ui.adsabs.harvard.edu/abs/2026MNRAS.548ag645C}
}

@ARTICLE{Rodriguez2021,
       author = {{Rodriguez}, Facundo and {Montero-Dorta}, Antonio D. and {Angulo}, Raul E. and {Artale}, M. Celeste and {Merch{\'a}n}, Manuel},
        title = "{The galaxy size-halo mass scaling relations and clustering properties of central and satellite galaxies}",
      journal = {\mnras},
         year = 2021,
        month = aug,
       volume = {505},
       number = {3},
        pages = {3192-3205},
          doi = {10.1093/mnras/stab1571},
archivePrefix = {arXiv},
       eprint = {2011.00014},
 primaryClass = {astro-ph.GA},
       adsurl = {https://ui.adsabs.harvard.edu/abs/2021MNRAS.505.3192R}
}

@ARTICLE{Hardwick2022,
       author = {{Hardwick}, Jennifer A. and {Cortese}, Luca and {Obreschkow}, Danail and {Catinella}, Barbara and {Cook}, Robin H.~W.},
        title = "{xGASS: characterizing the slope and scatter of the stellar mass-angular momentum relation for nearby galaxies}",
      journal = {\mnras},
         year = 2022,
        month = jan,
       volume = {509},
       number = {3},
        pages = {3751-3763},
          doi = {10.1093/mnras/stab3261},
archivePrefix = {arXiv},
       eprint = {2111.15048},
 primaryClass = {astro-ph.GA},
       adsurl = {https://ui.adsabs.harvard.edu/abs/2022MNRAS.509.3751H}
}

@ARTICLE{Wolf2010,
       author = {{Wolf}, Joe and {Martinez}, Gregory D. and {Bullock}, James S. and {Kaplinghat}, Manoj and {Geha}, Marla and {Mu{\~n}oz}, Ricardo R. and {Simon}, Joshua D. and {Avedo}, Frank F.},
        title = "{Accurate masses for dispersion-supported galaxies}",
      journal = {\mnras},
         year = 2010,
        month = aug,
       volume = {406},
       number = {2},
        pages = {1220-1237},
          doi = {10.1111/j.1365-2966.2010.16753.x},
archivePrefix = {arXiv},
       eprint = {0908.2995},
 primaryClass = {astro-ph.CO},
       adsurl = {https://ui.adsabs.harvard.edu/abs/2010MNRAS.406.1220W}
}

@ARTICLE{Kirby2013,
       author = {{Kirby}, Evan N. and {Cohen}, Judith G. and {Guhathakurta}, Puragra and {Cheng}, Lucy and {Bullock}, James S. and {Gallazzi}, Anna},
        title = "{The Universal Stellar Mass-Stellar Metallicity Relation for Dwarf Galaxies}",
      journal = {\apj},
         year = 2013,
        month = dec,
       volume = {779},
       number = {2},
          eid = {102},
        pages = {102},
          doi = {10.1088/0004-637X/779/2/102},
archivePrefix = {arXiv},
       eprint = {1310.0814},
 primaryClass = {astro-ph.GA},
       adsurl = {https://ui.adsabs.harvard.edu/abs/2013ApJ...779..102K}
}

@ARTICLE{Asali2025,
       author = {{Asali}, Yasmeen and {Geha}, Marla and {Kado-Fong}, Erin and {Mao}, Yao-Yuan and {Wechsler}, Risa H. and {de los Reyes}, Mithi A.~C. and {Pasha}, Imad and {Kallivayalil}, Nitya and {Nadler}, Ethan O. and {Tollerud}, Erik J. and {Wang}, Yunchong and {Weiner}, Benjamin and {Wu}, John F.},
        title = "{The SAGA Survey. VI. The Size─Mass Relation for Low-mass Galaxies Across Environments}",
      journal = {\apj},
         year = 2025,
        month = dec,
       volume = {995},
       number = {1},
          eid = {79},
        pages = {79},
          doi = {10.3847/1538-4357/ae147d},
archivePrefix = {arXiv},
       eprint = {2509.25335},
 primaryClass = {astro-ph.GA},
       adsurl = {https://ui.adsabs.harvard.edu/abs/2025ApJ...995...79A}
}

@ARTICLE{Oman2015,
       author = {{Oman}, Kyle A. and {Navarro}, Julio F. and {Fattahi}, Azadeh and {Frenk}, Carlos S. and {Sawala}, Till and {White}, Simon D.~M. and {Bower}, Richard and {Crain}, Robert A. and {Furlong}, Michelle and {Schaller}, Matthieu and {Schaye}, Joop and {Theuns}, Tom},
        title = "{The unexpected diversity of dwarf galaxy rotation curves}",
      journal = {\mnras},
         year = 2015,
        month = oct,
       volume = {452},
       number = {4},
        pages = {3650-3665},
          doi = {10.1093/mnras/stv1504},
archivePrefix = {arXiv},
       eprint = {1504.01437},
 primaryClass = {astro-ph.GA},
       adsurl = {https://ui.adsabs.harvard.edu/abs/2015MNRAS.452.3650O}
}

@ARTICLE{Fattahi2018,
       author = {{Fattahi}, Azadeh and {Navarro}, Julio F. and {Frenk}, Carlos S. and {Oman}, Kyle A. and {Sawala}, Till and {Schaller}, Matthieu},
        title = "{Tidal stripping and the structure of dwarf galaxies in the Local Group}",
      journal = {\mnras},
         year = 2018,
        month = may,
       volume = {476},
       number = {3},
        pages = {3816-3836},
          doi = {10.1093/mnras/sty408},
archivePrefix = {arXiv},
       eprint = {1707.03898},
 primaryClass = {astro-ph.GA},
       adsurl = {https://ui.adsabs.harvard.edu/abs/2018MNRAS.476.3816F}
}

@ARTICLE{Frenk1988,
       author = {{Frenk}, Carlos S. and {White}, Simon D.~M. and {Davis}, Marc and {Efstathiou}, George},
        title = "{The Formation of Dark Halos in a Universe Dominated by Cold Dark Matter}",
      journal = {\apj},
         year = 1988,
        month = apr,
       volume = {327},
        pages = {507},
          doi = {10.1086/166213},
       adsurl = {https://ui.adsabs.harvard.edu/abs/1988ApJ...327..507F}
}

@ARTICLE{Navarro1997,
       author = {{Navarro}, Julio F. and {Frenk}, Carlos S. and {White}, Simon D.~M.},
        title = "{A Universal Density Profile from Hierarchical Clustering}",
      journal = {\apj},
         year = 1997,
        month = dec,
       volume = {490},
       number = {2},
        pages = {493-508},
          doi = {10.1086/304888},
archivePrefix = {arXiv},
       eprint = {astro-ph/9611107},
 primaryClass = {astro-ph},
       adsurl = {https://ui.adsabs.harvard.edu/abs/1997ApJ...490..493N}
}

@ARTICLE{ReesOstriker1977,
       author = {{Rees}, M.~J. and {Ostriker}, J.~P.},
        title = "{Cooling, dynamics and fragmentation of massive gas clouds: clues to the masses and radii of galaxies and clusters.}",
      journal = {\mnras},
         year = 1977,
        month = jun,
       volume = {179},
        pages = {541-559},
          doi = {10.1093/mnras/179.4.541},
       adsurl = {https://ui.adsabs.harvard.edu/abs/1977MNRAS.179..541R}
}

@ARTICLE{FallEfstathiou1980,
       author = {{Fall}, S.~M. and {Efstathiou}, G.},
        title = "{Formation and rotation of disc galaxies with haloes.}",
      journal = {\mnras},
         year = 1980,
        month = oct,
       volume = {193},
        pages = {189-206},
          doi = {10.1093/mnras/193.2.189},
       adsurl = {https://ui.adsabs.harvard.edu/abs/1980MNRAS.193..189F}
}

@ARTICLE{Mo1998,
       author = {{Mo}, H.~J. and {Mao}, Shude and {White}, Simon D.~M.},
        title = "{The formation of galactic discs}",
      journal = {\mnras},
         year = 1998,
        month = apr,
       volume = {295},
       number = {2},
        pages = {319-336},
          doi = {10.1046/j.1365-8711.1998.01227.x},
archivePrefix = {arXiv},
       eprint = {astro-ph/9707093},
 primaryClass = {astro-ph},
       adsurl = {https://ui.adsabs.harvard.edu/abs/1998MNRAS.295..319M}
}

@ARTICLE{WechslerTinker2018,
       author = {{Wechsler}, Risa H. and {Tinker}, Jeremy L.},
        title = "{The Connection Between Galaxies and Their Dark Matter Halos}",
      journal = {\araa},
         year = 2018,
        month = sep,
       volume = {56},
        pages = {435-487},
          doi = {10.1146/annurev-astro-081817-051756},
archivePrefix = {arXiv},
       eprint = {1804.03097},
 primaryClass = {astro-ph.GA},
       adsurl = {https://ui.adsabs.harvard.edu/abs/2018ARA&A..56..435W}
}

@ARTICLE{DiCintio2014,
       author = {{Di Cintio}, Arianna and {Brook}, Chris B. and {Dutton}, Aaron A. and {Macci{\`o}}, Andrea V. and {Stinson}, Greg S. and {Knebe}, Alexander},
        title = "{A mass-dependent density profile for dark matter haloes including the influence of galaxy formation}",
      journal = {\mnras},
         year = 2014,
        month = jul,
       volume = {441},
       number = {4},
        pages = {2986-2995},
          doi = {10.1093/mnras/stu729},
archivePrefix = {arXiv},
       eprint = {1404.5959},
 primaryClass = {astro-ph.CO},
       adsurl = {https://ui.adsabs.harvard.edu/abs/2014MNRAS.441.2986D}
}

@ARTICLE{Oh2011,
       author = {{Oh}, Se-Heon and {Brook}, Chris and {Governato}, Fabio and {Brinks}, Elias and {Mayer}, Lucio and {de Blok}, W.~J.~G. and {Brooks}, Alyson and {Walter}, Fabian},
        title = "{The Central Slope of Dark Matter Cores in Dwarf Galaxies: Simulations versus THINGS}",
      journal = {\aj},
         year = 2011,
        month = jul,
       volume = {142},
       number = {1},
          eid = {24},
        pages = {24},
          doi = {10.1088/0004-6256/142/1/24},
archivePrefix = {arXiv},
       eprint = {1011.2777},
 primaryClass = {astro-ph.CO},
       adsurl = {https://ui.adsabs.harvard.edu/abs/2011AJ....142...24O}
}

@ARTICLE{Cruz2025,
       author = {{Cruz}, Akaxia and {Brooks}, Alyson and {Lisanti}, Mariangela and {Peter}, Annika H.~G. and {Geda}, Robel and {Quinn}, Thomas and {Tremmel}, Michael and {Munshi}, Ferah and {Keller}, Ben and {Wadsley}, James},
        title = "{Dwarf diversity in $Λ$CDM with baryons}",
      journal = {arXiv e-prints},
         year = 2025,
        month = oct,
          eid = {arXiv:2510.11800},
        pages = {arXiv:2510.11800},
          doi = {10.48550/arXiv.2510.11800},
archivePrefix = {arXiv},
       eprint = {2510.11800},
 primaryClass = {astro-ph.GA},
       adsurl = {https://ui.adsabs.harvard.edu/abs/2025arXiv251011800C}
}

@ARTICLE{Walker2009,
       author = {{Walker}, Matthew G. and {Mateo}, Mario and {Olszewski}, Edward W. and {Pe{\~n}arrubia}, Jorge and {Evans}, N. Wyn and {Gilmore}, Gerard},
        title = "{A Universal Mass Profile for Dwarf Spheroidal Galaxies?}",
      journal = {\apj},
         year = 2009,
        month = oct,
       volume = {704},
       number = {2},
        pages = {1274-1287},
          doi = {10.1088/0004-637X/704/2/1274},
archivePrefix = {arXiv},
       eprint = {0906.0341},
 primaryClass = {astro-ph.CO},
       adsurl = {https://ui.adsabs.harvard.edu/abs/2009ApJ...704.1274W}
}

@ARTICLE{Peng2015,
       author = {{Peng}, Y. and {Maiolino}, R. and {Cochrane}, R.},
        title = "{Strangulation as the primary mechanism for shutting down star formation in galaxies}",
      journal = {\nat},
         year = 2015,
        month = may,
       volume = {521},
       number = {7551},
        pages = {192-195},
          doi = {10.1038/nature14439},
archivePrefix = {arXiv},
       eprint = {1505.03143},
 primaryClass = {astro-ph.GA},
       adsurl = {https://ui.adsabs.harvard.edu/abs/2015Natur.521..192P}
}

@ARTICLE{GunnGott1972,
       author = {{Gunn}, James E. and {Gott}, III, J. Richard},
        title = "{On the Infall of Matter Into Clusters of Galaxies and Some Effects on Their Evolution}",
      journal = {\apj},
         year = 1972,
        month = aug,
       volume = {176},
        pages = {1},
          doi = {10.1086/151605},
       adsurl = {https://ui.adsabs.harvard.edu/abs/1972ApJ...176....1G}
}

@ARTICLE{Moore1996Natur,
       author = {{Moore}, Ben and {Katz}, Neal and {Lake}, George and {Dressler}, Alan and {Oemler}, Augustus},
        title = "{Galaxy harassment and the evolution of clusters of galaxies}",
      journal = {\nat},
         year = 1996,
        month = feb,
       volume = {379},
       number = {6566},
        pages = {613-616},
          doi = {10.1038/379613a0},
archivePrefix = {arXiv},
       eprint = {astro-ph/9510034},
 primaryClass = {astro-ph},
       adsurl = {https://ui.adsabs.harvard.edu/abs/1996Natur.379..613M}
}

@ARTICLE{Moore1998,
       author = {{Moore}, Ben and {Lake}, George and {Katz}, Neal},
        title = "{Morphological Transformation from Galaxy Harassment}",
      journal = {\apj},
         year = 1998,
        month = mar,
       volume = {495},
       number = {1},
        pages = {139-151},
          doi = {10.1086/305264},
archivePrefix = {arXiv},
       eprint = {astro-ph/9701211},
 primaryClass = {astro-ph},
       adsurl = {https://ui.adsabs.harvard.edu/abs/1998ApJ...495..139M}
}

@ARTICLE{TaylorBabul2004,
       author = {{Taylor}, James E. and {Babul}, Arif},
        title = "{The evolution of substructure in galaxy, group and cluster haloes - I. Basic dynamics}",
      journal = {\mnras},
         year = 2004,
        month = mar,
       volume = {348},
       number = {3},
        pages = {811-830},
          doi = {10.1111/j.1365-2966.2004.07395.x},
archivePrefix = {arXiv},
       eprint = {astro-ph/0301612},
 primaryClass = {astro-ph},
       adsurl = {https://ui.adsabs.harvard.edu/abs/2004MNRAS.348..811T}
}

@ARTICLE{Kauffmann2004,
       author = {{Kauffmann}, Guinevere and {White}, Simon D.~M. and {Heckman}, Timothy M. and {M{\'e}nard}, Brice and {Brinchmann}, Jarle and {Charlot}, St{\'e}phane and {Tremonti}, Christy and {Brinkmann}, Jon},
        title = "{The environmental dependence of the relations between stellar mass, structure, star formation and nuclear activity in galaxies}",
      journal = {\mnras},
         year = 2004,
        month = sep,
       volume = {353},
       number = {3},
        pages = {713-731},
          doi = {10.1111/j.1365-2966.2004.08117.x},
archivePrefix = {arXiv},
       eprint = {astro-ph/0402030},
 primaryClass = {astro-ph},
       adsurl = {https://ui.adsabs.harvard.edu/abs/2004MNRAS.353..713K}
}

@ARTICLE{Wetzel2013,
       author = {{Wetzel}, Andrew R. and {Tinker}, Jeremy L. and {Conroy}, Charlie and {van den Bosch}, Frank C.},
        title = "{Galaxy evolution in groups and clusters: satellite star formation histories and quenching time-scales in a hierarchical Universe}",
      journal = {\mnras},
         year = 2013,
        month = jun,
       volume = {432},
       number = {1},
        pages = {336-358},
          doi = {10.1093/mnras/stt469},
archivePrefix = {arXiv},
       eprint = {1206.3571},
 primaryClass = {astro-ph.CO},
       adsurl = {https://ui.adsabs.harvard.edu/abs/2013MNRAS.432..336W}
}

@ARTICLE{Peng2012,
       author = {{Peng}, Ying-jie and {Lilly}, Simon J. and {Renzini}, Alvio and {Carollo}, Marcella},
        title = "{Mass and Environment as Drivers of Galaxy Evolution. II. The Quenching of Satellite Galaxies as the Origin of Environmental Effects}",
      journal = {\apj},
         year = 2012,
        month = sep,
       volume = {757},
       number = {1},
          eid = {4},
        pages = {4},
          doi = {10.1088/0004-637X/757/1/4},
archivePrefix = {arXiv},
       eprint = {1106.2546},
 primaryClass = {astro-ph.CO},
       adsurl = {https://ui.adsabs.harvard.edu/abs/2012ApJ...757....4P}
}

@ARTICLE{Merritt1983,
       author = {{Merritt}, D.},
        title = "{Relaxation and tidal stripping in rich clusters of galaxies. I. Evolution of the mass distribution.}",
      journal = {\apj},
         year = 1983,
        month = jan,
       volume = {264},
        pages = {24-48},
          doi = {10.1086/160571},
       adsurl = {https://ui.adsabs.harvard.edu/abs/1983ApJ...264...24M}
}

@ARTICLE{Buck2019,
       author = {{Buck}, Tobias and {Macci{\`o}}, Andrea V. and {Dutton}, Aaron A. and {Obreja}, Aura and {Frings}, Jonas},
        title = "{NIHAO XV: the environmental impact of the host galaxy on galactic satellite and field dwarf galaxies}",
      journal = {\mnras},
         year = 2019,
        month = feb,
       volume = {483},
       number = {1},
        pages = {1314-1341},
          doi = {10.1093/mnras/sty2913},
archivePrefix = {arXiv},
       eprint = {1804.04667},
 primaryClass = {astro-ph.GA},
       adsurl = {https://ui.adsabs.harvard.edu/abs/2019MNRAS.483.1314B}
}

@ARTICLE{Chaikin2026,
       author = {{Chaikin}, Evgenii and {Schaye}, Joop and {Schaller}, Matthieu and {Ploeckinger}, Sylvia and {Bah{\'e}}, Yannick M. and {Ben{\'\i}tez-Llambay}, Alejandro and {Correa}, Camila and {Forouhar Moreno}, Victor J. and {Frenk}, Carlos S. and {Hu{\v{s}}ko}, Filip and {Kugel}, Roi and {McGibbon}, Robert and {Richings}, Alexander J. and {Trayford}, James W. and {Borrow}, Josh and {Crain}, Robert A. and {Helly}, John C. and {Lacey}, Cedric G. and {Ludlow}, Aaron and {Nobels}, Folkert S.~J.},
        title = "{COLIBRE: calibrating subgrid feedback in cosmological simulations that include a cold gas phase}",
      journal = {\mnras},
         year = 2026,
        month = may,
       volume = {548},
       number = {1},
          eid = {stag300},
        pages = {stag300},
          doi = {10.1093/mnras/stag300},
archivePrefix = {arXiv},
       eprint = {2509.04067},
 primaryClass = {astro-ph.GA},
       adsurl = {https://ui.adsabs.harvard.edu/abs/2026MNRAS.548ag300C}
}

@ARTICLE{Schaye2015,
       author = {{Schaye}, Joop and {Crain}, Robert A. and {Bower}, Richard G. and {Furlong}, Michelle and {Schaller}, Matthieu and {Theuns}, Tom and {Dalla Vecchia}, Claudio and {Frenk}, Carlos S. and {McCarthy}, I.~G. and {Helly}, John C. and {Jenkins}, Adrian and {Rosas-Guevara}, Y.~M. and {White}, Simon D.~M. and {Baes}, Maarten and {Booth}, C.~M. and {Camps}, Peter and {Navarro}, Julio F. and {Qu}, Yan and {Rahmati}, Alireza and {Sawala}, Till and {Thomas}, Peter A. and {Trayford}, James},
        title = "{The EAGLE project: simulating the evolution and assembly of galaxies and their environments}",
      journal = {\mnras},
         year = 2015,
        month = jan,
       volume = {446},
       number = {1},
        pages = {521-554},
          doi = {10.1093/mnras/stu2058},
archivePrefix = {arXiv},
       eprint = {1407.7040},
 primaryClass = {astro-ph.GA},
       adsurl = {https://ui.adsabs.harvard.edu/abs/2015MNRAS.446..521S}
}

@ARTICLE{Crain2015,
       author = {{Crain}, Robert A. and {Schaye}, Joop and {Bower}, Richard G. and {Furlong}, Michelle and {Schaller}, Matthieu and {Theuns}, Tom and {Dalla Vecchia}, Claudio and {Frenk}, Carlos S. and {McCarthy}, Ian G. and {Helly}, John C. and {Jenkins}, Adrian and {Rosas-Guevara}, Yetli M. and {White}, Simon D.~M. and {Trayford}, James W.},
        title = "{The EAGLE simulations of galaxy formation: calibration of subgrid physics and model variations}",
      journal = {\mnras},
         year = 2015,
        month = jun,
       volume = {450},
       number = {2},
        pages = {1937-1961},
          doi = {10.1093/mnras/stv725},
archivePrefix = {arXiv},
       eprint = {1501.01311},
 primaryClass = {astro-ph.GA},
       adsurl = {https://ui.adsabs.harvard.edu/abs/2015MNRAS.450.1937C}
}

@ARTICLE{Abbott2022,
       author = {{Abbott}, T.~M.~C. and {Aguena}, M. and {Alarcon}, A. and {Allam}, S. and {Alves}, O. and {Amon}, A. and {Andrade-Oliveira}, F. and {Annis}, J. and {Avila}, S. and {Bacon}, D. and {Baxter}, E. and {Bechtol}, K. and {Becker}, M.~R. and {Bernstein}, G.~M. and {Bhargava}, S. and {Birrer}, S. and {Blazek}, J. and {Brandao-Souza}, A. and {Bridle}, S.~L. and {Brooks}, D. and {Buckley-Geer}, E. and {Burke}, D.~L. and {Camacho}, H. and {Campos}, A. and {Carnero Rosell}, A. and {Carrasco Kind}, M. and {Carretero}, J. and {Castander}, F.~J. and {Cawthon}, R. and {Chang}, C. and {Chen}, A. and {Chen}, R. and {Choi}, A. and {Conselice}, C. and {Cordero}, J. and {Costanzi}, M. and {Crocce}, M. and {da Costa}, L.~N. and {da Silva Pereira}, M.~E. and {Davis}, C. and {Davis}, T.~M. and {De Vicente}, J. and {DeRose}, J. and {Desai}, S. and {Di Valentino}, E. and {Diehl}, H.~T. and {Dietrich}, J.~P. and {Dodelson}, S. and {Doel}, P. and {Doux}, C. and {Drlica-Wagner}, A. and {Eckert}, K. and {Eifler}, T.~F. and {Elsner}, F. and {Elvin-Poole}, J. and {Everett}, S. and {Evrard}, A.~E. and {Fang}, X. and {Farahi}, A. and {Fernandez}, E. and {Ferrero}, I. and {Fert{\'e}}, A. and {Fosalba}, P. and {Friedrich}, O. and {Frieman}, J. and {Garc{\'\i}a-Bellido}, J. and {Gatti}, M. and {Gaztanaga}, E. and {Gerdes}, D.~W. and {Giannantonio}, T. and {Giannini}, G. and {Gruen}, D. and {Gruendl}, R.~A. and {Gschwend}, J. and {Gutierrez}, G. and {Harrison}, I. and {Hartley}, W.~G. and {Herner}, K. and {Hinton}, S.~R. and {Hollowood}, D.~L. and {Honscheid}, K. and {Hoyle}, B. and {Huff}, E.~M. and {Huterer}, D. and {Jain}, B. and {James}, D.~J. and {Jarvis}, M. and {Jeffrey}, N. and {Jeltema}, T. and {Kovacs}, A. and {Krause}, E. and {Kron}, R. and {Kuehn}, K. and {Kuropatkin}, N. and {Lahav}, O. and {Leget}, P.-F. and {Lemos}, P. and {Liddle}, A.~R. and {Lidman}, C. and {Lima}, M. and {Lin}, H. and {MacCrann}, N. and {Maia}, M.~A.~G. and {Marshall}, J.~L. and {Martini}, P. and {McCullough}, J. and {Melchior}, P. and {Mena-Fern{\'a}ndez}, J. and {Menanteau}, F. and {Miquel}, R. and {Mohr}, J.~J. and {Morgan}, R. and {Muir}, J. and {Myles}, J. and {Nadathur}, S. and {Navarro-Alsina}, A. and {Nichol}, R.~C. and {Ogando}, R.~L.~C. and {Omori}, Y. and {Palmese}, A. and {Pandey}, S. and {Park}, Y. and {Paz-Chinch{\'o}n}, F. and {Petravick}, D. and {Pieres}, A. and {Plazas Malag{\'o}n}, A.~A. and {Porredon}, A. and {Prat}, J. and {Raveri}, M. and {Rodriguez-Monroy}, M. and {Rollins}, R.~P. and {Romer}, A.~K. and {Roodman}, A. and {Rosenfeld}, R. and {Ross}, A.~J. and {Rykoff}, E.~S. and {Samuroff}, S. and {S{\'a}nchez}, C. and {Sanchez}, E. and {Sanchez}, J. and {Sanchez Cid}, D. and {Scarpine}, V. and {Schubnell}, M. and {Scolnic}, D. and {Secco}, L.~F. and {Serrano}, S. and {Sevilla-Noarbe}, I. and {Sheldon}, E. and {Shin}, T. and {Smith}, M. and {Soares-Santos}, M. and {Suchyta}, E. and {Swanson}, M.~E.~C. and {Tabbutt}, M. and {Tarle}, G. and {Thomas}, D. and {To}, C. and {Troja}, A. and {Troxel}, M.~A. and {Tucker}, D.~L. and {Tutusaus}, I. and {Varga}, T.~N. and {Walker}, A.~R. and {Weaverdyck}, N. and {Wechsler}, R. and {Weller}, J. and {Yanny}, B. and {Yin}, B. and {Zhang}, Y. and {Zuntz}, J. and {DES Collaboration}},
        title = "{Dark Energy Survey Year 3 results: Cosmological constraints from galaxy clustering and weak lensing}",
      journal = {\prd},
         year = 2022,
        month = jan,
       volume = {105},
       number = {2},
          eid = {023520},
        pages = {023520},
          doi = {10.1103/PhysRevD.105.023520},
archivePrefix = {arXiv},
       eprint = {2105.13549},
 primaryClass = {astro-ph.CO},
       adsurl = {https://ui.adsabs.harvard.edu/abs/2022PhRvD.105b3520A}
}

@ARTICLE{Schaller2024,
       author = {{Schaller}, Matthieu and {Borrow}, Josh and {Draper}, Peter W. and {Ivkovic}, Mladen and {McAlpine}, Stuart and {Vandenbroucke}, Bert and {Bah{\'e}}, Yannick and {Chaikin}, Evgenii and {Chalk}, Aidan B.~G. and {Chan}, Tsang Keung and {Correa}, Camila and {van Daalen}, Marcel and {Elbers}, Willem and {Gonnet}, Pedro and {Hausammann}, Lo{\"\i}c and {Helly}, John and {Hu{\v{s}}ko}, Filip and {Kegerreis}, Jacob A. and {Nobels}, Folkert S.~J. and {Ploeckinger}, Sylvia and {Revaz}, Yves and {Roper}, William J. and {Ruiz-Bonilla}, Sergio and {Sandnes}, Thomas D. and {Uyttenhove}, Yolan and {Willis}, James S. and {Xiang}, Zhen},
        title = "{SWIFT: A modern highly-parallel gravity and smoothed particle hydrodynamics solver for astrophysical and cosmological applications}",
      journal = {\mnras},
         year = 2024,
        month = may,
       volume = {530},
       number = {2},
        pages = {2378-2419},
          doi = {10.1093/mnras/stae922},
archivePrefix = {arXiv},
       eprint = {2305.13380},
 primaryClass = {astro-ph.IM},
       adsurl = {https://ui.adsabs.harvard.edu/abs/2024MNRAS.530.2378S}
}

@ARTICLE{Forouhar2025,
       author = {{Forouhar Moreno}, Victor J. and {Helly}, John and {McGibbon}, Robert and {Schaye}, Joop and {Schaller}, Matthieu and {Han}, Jiaxin and {Kugel}, Roi and {Bah{\'e}}, Yannick M.},
        title = "{Assessing subhalo finders in cosmological hydrodynamical simulations}",
      journal = {\mnras},
         year = 2025,
        month = oct,
       volume = {543},
       number = {2},
        pages = {1339-1372},
          doi = {10.1093/mnras/staf1478},
archivePrefix = {arXiv},
       eprint = {2502.06932},
 primaryClass = {astro-ph.CO},
       adsurl = {https://ui.adsabs.harvard.edu/abs/2025MNRAS.543.1339F}
}

@ARTICLE{Han2018,
       author = {{Han}, Jiaxin and {Cole}, Shaun and {Frenk}, Carlos S. and {Benitez-Llambay}, Alejandro and {Helly}, John},
        title = "{HBT+: an improved code for finding subhaloes and building merger trees in cosmological simulations}",
      journal = {\mnras},
         year = 2018,
        month = feb,
       volume = {474},
       number = {1},
        pages = {604-617},
          doi = {10.1093/mnras/stx2792},
archivePrefix = {arXiv},
       eprint = {1708.03646},
 primaryClass = {astro-ph.CO},
       adsurl = {https://ui.adsabs.harvard.edu/abs/2018MNRAS.474..604H}
}

@ARTICLE{McGibbon2025,
       author = {{McGibbon}, Robert and {Helly}, John and {Schaye}, Joop and {Schaller}, Matthieu and {Vandenbroucke}, Bert},
        title = "{SOAP: A Python Package for Calculating the Properties of Galaxies and Halos Formed in Cosmological Simulations}",
      journal = {The Journal of Open Source Software},
         year = 2025,
        month = jul,
       volume = {10},
       number = {111},
          eid = {8252},
        pages = {8252},
          doi = {10.21105/joss.08252},
archivePrefix = {arXiv},
       eprint = {2507.22669},
 primaryClass = {astro-ph.IM},
       adsurl = {https://ui.adsabs.harvard.edu/abs/2025JOSS...10.8252M}
}

@ARTICLE{deGraaff2022,
       author = {{de Graaff}, Anna and {Trayford}, James and {Franx}, Marijn and {Schaller}, Matthieu and {Schaye}, Joop and {van der Wel}, Arjen},
        title = "{Observed structural parameters of EAGLE galaxies: reconciling the mass-size relation in simulations with local observations}",
      journal = {\mnras},
         year = 2022,
        month = apr,
       volume = {511},
       number = {2},
        pages = {2544-2564},
          doi = {10.1093/mnras/stab3510},
archivePrefix = {arXiv},
       eprint = {2110.02235},
 primaryClass = {astro-ph.GA},
       adsurl = {https://ui.adsabs.harvard.edu/abs/2022MNRAS.511.2544D}
}

@ARTICLE{Nobels2024,
       author = {{Nobels}, Folkert S.~J. and {Schaye}, Joop and {Schaller}, Matthieu and {Ploeckinger}, Sylvia and {Chaikin}, Evgenii and {Richings}, Alexander J.},
        title = "{Tests of subgrid models for star formation using simulations of isolated disc galaxies}",
      journal = {\mnras},
         year = 2024,
        month = aug,
       volume = {532},
       number = {3},
        pages = {3299-3321},
          doi = {10.1093/mnras/stae1390},
archivePrefix = {arXiv},
       eprint = {2309.13750},
 primaryClass = {astro-ph.GA},
       adsurl = {https://ui.adsabs.harvard.edu/abs/2024MNRAS.532.3299N}
}

@ARTICLE{Ploeckinger2025,
       author = {{Ploeckinger}, Sylvia and {Richings}, Alexander J. and {Schaye}, Joop and {Trayford}, James W. and {Schaller}, Matthieu and {Chaikin}, Evgenii},
        title = "{HYBRID-CHIMES: a model for radiative cooling and the abundances of ions and molecules in simulations of galaxy formation}",
      journal = {\mnras},
         year = 2025,
        month = oct,
       volume = {543},
       number = {2},
        pages = {891-916},
          doi = {10.1093/mnras/staf1402},
archivePrefix = {arXiv},
       eprint = {2506.15773},
 primaryClass = {astro-ph.GA},
       adsurl = {https://ui.adsabs.harvard.edu/abs/2025MNRAS.543..891P}
}

@ARTICLE{Trayford2026,
       author = {{Trayford}, James W. and {Schaye}, Joop and {Correa}, Camila and {Ploeckinger}, Sylvia and {Richings}, Alexander J. and {Chaikin}, Evgenii and {Schaller}, Matthieu and {Ben{\'\i}tez-Llambay}, Alejandro and {Frenk}, Carlos and {Hu{\v{s}}ko}, Filip},
        title = "{Modelling the evolution and influence of dust in cosmological simulations that include the cold phase of the interstellar medium}",
      journal = {\mnras},
         year = 2026,
        month = feb,
       volume = {545},
       number = {4},
          eid = {staf2040},
        pages = {staf2040},
          doi = {10.1093/mnras/staf2040},
archivePrefix = {arXiv},
       eprint = {2505.13056},
 primaryClass = {astro-ph.GA},
       adsurl = {https://ui.adsabs.harvard.edu/abs/2026MNRAS.545f2040T}
}

@ARTICLE{Husko2026,
       author = {{Hu{\v{s}}ko}, Filip and {Lacey}, Cedric G. and {Schaye}, Joop and {Schaller}, Matthieu and {Chaikin}, Evgenii and {Ploeckinger}, Sylvia and {Ben{\'\i}tez Llambay}, Alejandro and {Richings}, Alexander J. and {Trayford}, James W.},
        title = "{A hybrid active galactic nucleus feedback model with spinning black holes, winds and jets}",
      journal = {\mnras},
         year = 2026,
        month = apr,
       volume = {547},
       number = {2},
          eid = {stag324},
        pages = {stag324},
          doi = {10.1093/mnras/stag324},
archivePrefix = {arXiv},
       eprint = {2509.05179},
 primaryClass = {astro-ph.GA},
       adsurl = {https://ui.adsabs.harvard.edu/abs/2026MNRAS.547ag324H}
}

@ARTICLE{BL2026,
       author = {{Ben{\'\i}tez-Llambay}, Alejandro and {Ploeckinger}, Sylvia and {Schaye}, Joop and {Richings}, Alexander J. and {Chaikin}, Evgenii and {Schaller}, Matthieu and {Trayford}, James W. and {Frenk}, Carlos S. and {Hu{\v{s}}ko}, Filip and {Correa}, Camila},
        title = "{Non-explosive pre-supernova feedback in the COLIBRE model of galaxy formation}",
      journal = {\mnras},
         year = 2026,
        month = mar,
       volume = {546},
       number = {4},
          eid = {stag268},
        pages = {stag268},
          doi = {10.1093/mnras/stag268},
archivePrefix = {arXiv},
       eprint = {2509.25309},
 primaryClass = {astro-ph.GA},
       adsurl = {https://ui.adsabs.harvard.edu/abs/2026MNRAS.546ag268B}
}

@ARTICLE{Chaikin2023,
       author = {{Chaikin}, Evgenii and {Schaye}, Joop and {Schaller}, Matthieu and {Ben{\'\i}tez-Llambay}, Alejandro and {Nobels}, Folkert S.~J. and {Ploeckinger}, Sylvia},
        title = "{A thermal-kinetic subgrid model for supernova feedback in simulations of galaxy formation}",
      journal = {\mnras},
         year = 2023,
        month = aug,
       volume = {523},
       number = {3},
        pages = {3709-3731},
          doi = {10.1093/mnras/stad1626},
archivePrefix = {arXiv},
       eprint = {2211.04619},
 primaryClass = {astro-ph.GA},
       adsurl = {https://ui.adsabs.harvard.edu/abs/2023MNRAS.523.3709C}
}

@ARTICLE{Errani2024,
       author = {{Errani}, Rapha{\"e}l and {Ibata}, Rodrigo and {Navarro}, Julio F. and {Pe{\~n}arrubia}, Jorge and {Walker}, Matthew G.},
        title = "{Microgalaxies in LCDM}",
      journal = {\apj},
         year = 2024,
        month = jun,
       volume = {968},
       number = {2},
          eid = {89},
        pages = {89},
          doi = {10.3847/1538-4357/ad402d},
archivePrefix = {arXiv},
       eprint = {2311.14798},
 primaryClass = {astro-ph.GA},
       adsurl = {https://ui.adsabs.harvard.edu/abs/2024ApJ...968...89E}
}

@ARTICLE{Chandrasekhar1943,
       author = {{Chandrasekhar}, S.},
        title = "{Dynamical Friction. I. General Considerations: the Coefficient of Dynamical Friction.}",
      journal = {\apj},
         year = 1943,
        month = mar,
       volume = {97},
        pages = {255},
          doi = {10.1086/144517},
       adsurl = {https://ui.adsabs.harvard.edu/abs/1943ApJ....97..255C}
}

@ARTICLE{Matteucci1994,
       author = {{Matteucci}, Francesca},
        title = "{Abundance ratios in ellipticals and galaxy formation.}",
      journal = {\aap},
         year = 1994,
        month = aug,
       volume = {288},
        pages = {57-64},
       adsurl = {https://ui.adsabs.harvard.edu/abs/1994A&A...288...57M}
}

@ARTICLE{DekelSilk1986,
       author = {{Dekel}, A. and {Silk}, J.},
        title = "{The Origin of Dwarf Galaxies, Cold Dark Matter, and Biased Galaxy Formation}",
      journal = {\apj},
         year = 1986,
        month = apr,
       volume = {303},
        pages = {39},
          doi = {10.1086/164050},
       adsurl = {https://ui.adsabs.harvard.edu/abs/1986ApJ...303...39D}
}

@ARTICLE{Weisz2015,
       author = {{Weisz}, Daniel R. and {Dolphin}, Andrew E. and {Skillman}, Evan D. and {Holtzman}, Jon and {Gilbert}, Karoline M. and {Dalcanton}, Julianne J. and {Williams}, Benjamin F.},
        title = "{The Star Formation Histories of Local Group Dwarf Galaxies. III. Characterizing Quenching in Low-mass Galaxies}",
      journal = {\apj},
         year = 2015,
        month = may,
       volume = {804},
       number = {2},
          eid = {136},
        pages = {136},
          doi = {10.1088/0004-637X/804/2/136},
archivePrefix = {arXiv},
       eprint = {1503.05195},
 primaryClass = {astro-ph.GA},
       adsurl = {https://ui.adsabs.harvard.edu/abs/2015ApJ...804..136W}
}

@ARTICLE{Pasquali2012,
       author = {{Pasquali}, Anna and {Gallazzi}, Anna and {van den Bosch}, Frank C.},
        title = "{The gas-phase metallicity of central and satellite galaxies in the Sloan Digital Sky Survey}",
      journal = {\mnras},
         year = 2012,
        month = sep,
       volume = {425},
       number = {1},
        pages = {273-286},
          doi = {10.1111/j.1365-2966.2012.21454.x},
archivePrefix = {arXiv},
       eprint = {1206.3458},
 primaryClass = {astro-ph.CO},
       adsurl = {https://ui.adsabs.harvard.edu/abs/2012MNRAS.425..273P}
}

@ARTICLE{Abadi1999,
       author = {{Abadi}, Mario G. and {Moore}, Ben and {Bower}, Richard G.},
        title = "{Ram pressure stripping of spiral galaxies in clusters}",
      journal = {\mnras},
         year = 1999,
        month = oct,
       volume = {308},
       number = {4},
        pages = {947-954},
          doi = {10.1046/j.1365-8711.1999.02715.x},
archivePrefix = {arXiv},
       eprint = {astro-ph/9903436},
 primaryClass = {astro-ph},
       adsurl = {https://ui.adsabs.harvard.edu/abs/1999MNRAS.308..947A}
}

@ARTICLE{Gnedin1999,
       author = {{Gnedin}, Oleg Y. and {Lee}, Hyung Mok and {Ostriker}, Jeremiah P.},
        title = "{Effects of Tidal Shocks on the Evolution of Globular Clusters}",
      journal = {\apj},
         year = 1999,
        month = sep,
       volume = {522},
       number = {2},
        pages = {935-949},
          doi = {10.1086/307659},
archivePrefix = {arXiv},
       eprint = {astro-ph/9806245},
 primaryClass = {astro-ph},
       adsurl = {https://ui.adsabs.harvard.edu/abs/1999ApJ...522..935G}
}

@ARTICLE{Hernquist1990,
       author = {{Hernquist}, Lars},
        title = "{An Analytical Model for Spherical Galaxies and Bulges}",
      journal = {\apj},
         year = 1990,
        month = jun,
       volume = {356},
        pages = {359},
          doi = {10.1086/168845},
       adsurl = {https://ui.adsabs.harvard.edu/abs/1990ApJ...356..359H}
}

@ARTICLE{Lazar2020,
       author = {{Lazar}, Alexandres and {Bullock}, James S. and {Boylan-Kolchin}, Michael and {Chan}, T.~K. and {Hopkins}, Philip F. and {Graus}, Andrew S. and {Wetzel}, Andrew and {El-Badry}, Kareem and {Wheeler}, Coral and {Straight}, Maria C. and {Kere{\v{s}}}, Du{\v{s}}an and {Faucher-Gigu{\`e}re}, Claude-Andr{\'e} and {Fitts}, Alex and {Garrison-Kimmel}, Shea},
        title = "{A dark matter profile to model diverse feedback-induced core sizes of {\ensuremath{\Lambda}}CDM haloes}",
      journal = {\mnras},
         year = 2020,
        month = sep,
       volume = {497},
       number = {2},
        pages = {2393-2417},
          doi = {10.1093/mnras/staa2101},
archivePrefix = {arXiv},
       eprint = {2004.10817},
 primaryClass = {astro-ph.GA},
       adsurl = {https://ui.adsabs.harvard.edu/abs/2020MNRAS.497.2393L}
}

@article{Borrow2020, doi = {10.21105/joss.02430}, url = {https://doi.org/10.21105/joss.02430}, year = {2020}, publisher = {The Open Journal}, volume = {5}, number = {52}, pages = {2430}, author = {Borrow, Josh and Borrisov, Alexei}, title = {swiftsimio: A Python library for reading SWIFT data}, journal = {Journal of Open Source Software}}

@ARTICLE{Burkert1995,
       author = {{Burkert}, A.},
        title = "{The Structure of Dark Matter Halos in Dwarf Galaxies}",
      journal = {\apjl},
         year = 1995,
        month = jul,
       volume = {447},
        pages = {L25-L28},
          doi = {10.1086/309560},
archivePrefix = {arXiv},
       eprint = {astro-ph/9504041},
 primaryClass = {astro-ph},
       adsurl = {https://ui.adsabs.harvard.edu/abs/1995ApJ...447L..25B}
}

@ARTICLE{Wright2022,
       author = {{Wright}, Ruby J. and {Lagos}, Claudia del P. and {Power}, Chris and {Stevens}, Adam R.~H. and {Cortese}, Luca and {Poulton}, Rhys J.~J.},
        title = "{An orbital perspective on the starvation, stripping, and quenching of satellite galaxies in the EAGLE simulations}",
      journal = {\mnras},
         year = 2022,
        month = oct,
       volume = {516},
       number = {2},
        pages = {2891-2912},
          doi = {10.1093/mnras/stac2042},
archivePrefix = {arXiv},
       eprint = {2205.08414},
 primaryClass = {astro-ph.GA},
       adsurl = {https://ui.adsabs.harvard.edu/abs/2022MNRAS.516.2891W}
}

@ARTICLE{Gallazzi2021,
       author = {{Gallazzi}, Anna R. and {Pasquali}, A. and {Zibetti}, S. and {Barbera}, F. La},
        title = "{Galaxy evolution across environments as probed by the ages, stellar metallicities, and [{\ensuremath{\alpha}} /Fe] of central and satellite galaxies}",
      journal = {\mnras},
         year = 2021,
        month = apr,
       volume = {502},
       number = {3},
        pages = {4457-4478},
          doi = {10.1093/mnras/stab265},
archivePrefix = {arXiv},
       eprint = {2010.04733},
 primaryClass = {astro-ph.GA},
       adsurl = {https://ui.adsabs.harvard.edu/abs/2021MNRAS.502.4457G}
}

@ARTICLE{Wang2020,
       author = {{Wang}, Enci and {Wang}, Huiyuan and {Mo}, Houjun and {van den Bosch}, Frank C. and {Yang}, Xiaohu},
        title = "{The Dearth of Differences between Central and Satellite Galaxies. III. Environmental Dependencies of Mass-Size and Mass-Structure Relations}",
      journal = {\apj},
         year = 2020,
        month = jan,
       volume = {889},
       number = {1},
          eid = {37},
        pages = {37},
          doi = {10.3847/1538-4357/ab6217},
archivePrefix = {arXiv},
       eprint = {1912.05969},
 primaryClass = {astro-ph.GA},
       adsurl = {https://ui.adsabs.harvard.edu/abs/2020ApJ...889...37W}
}

@ARTICLE{Ludlow2021,
       author = {{Ludlow}, Aaron D. and {Fall}, S. Michael and {Schaye}, Joop and {Obreschkow}, Danail},
        title = "{Spurious heating of stellar motions in simulated galactic discs by dark matter halo particles}",
      journal = {\mnras},
         year = 2021,
        month = dec,
       volume = {508},
       number = {4},
        pages = {5114-5137},
          doi = {10.1093/mnras/stab2770},
archivePrefix = {arXiv},
       eprint = {2105.03561},
 primaryClass = {astro-ph.GA},
       adsurl = {https://ui.adsabs.harvard.edu/abs/2021MNRAS.508.5114L}
}

@ARTICLE{Celiz2025,
       author = {{Celiz}, Bruno M. and {Navarro}, Julio F. and {Abadi}, Mario G. and {Springel}, Volker},
        title = "{Mass-morphology relation of TNG50 galaxies}",
      journal = {\aap},
         year = 2025,
        month = jul,
       volume = {699},
          eid = {A12},
        pages = {A12},
          doi = {10.1051/0004-6361/202554847},
archivePrefix = {arXiv},
       eprint = {2505.01620},
 primaryClass = {astro-ph.GA},
       adsurl = {https://ui.adsabs.harvard.edu/abs/2025A&A...699A..12C}
}

@ARTICLE{Zeng2024,
       author = {{Zeng}, Guangquan and {Wang}, Lan and {Gao}, Liang and {Yang}, Hang},
        title = "{Kinematic morphology of low-mass galaxies in IllustrisTNG}",
      journal = {\mnras},
         year = 2024,
        month = aug,
       volume = {532},
       number = {2},
        pages = {2558-2569},
          doi = {10.1093/mnras/stae1651},
archivePrefix = {arXiv},
       eprint = {2404.14184},
 primaryClass = {astro-ph.GA},
       adsurl = {https://ui.adsabs.harvard.edu/abs/2024MNRAS.532.2558Z}
}

@ARTICLE{Abdullah2026,
       author = {{Abdullah}, Mohamed H. and {Abdelhamid}, Nouran E. and {Samir}, Rasha M. and {Wilson}, Gillian},
        title = "{From the Densest Clusters to the Emptiest Voids: No Evidence for Environmental Effects on the Galaxy Size─Stellar Mass Relation at Low Redshift}",
      journal = {\apj},
         year = 2026,
        month = may,
       volume = {1002},
       number = {2},
          eid = {173},
        pages = {173},
          doi = {10.3847/1538-4357/ae6113},
archivePrefix = {arXiv},
       eprint = {2507.04212},
 primaryClass = {astro-ph.GA},
       adsurl = {https://ui.adsabs.harvard.edu/abs/2026ApJ..1002..173A}
}

@ARTICLE{SantosSantos2025,
       author = {{Santos-Santos}, Isabel M.~E. and {Frenk}, Carlos S. and {Navarro}, Julio F. and {Cole}, Shaun and {Helly}, John},
        title = "{The unabridged satellite luminosity function of Milky Way-like galaxies in {\ensuremath{\Lambda}}CDM: the contribution of 'orphan' satellites}",
      journal = {\mnras},
         year = 2025,
        month = jun,
       volume = {540},
       number = {1},
        pages = {1107-1123},
          doi = {10.1093/mnras/staf749},
archivePrefix = {arXiv},
       eprint = {2410.19475},
 primaryClass = {astro-ph.GA},
       adsurl = {https://ui.adsabs.harvard.edu/abs/2025MNRAS.540.1107S}
}

@ARTICLE{vandenBosch2018,
       author = {{van den Bosch}, Frank C. and {Ogiya}, Go},
        title = "{Dark matter substructure in numerical simulations: a tale of discreteness noise, runaway instabilities, and artificial disruption}",
      journal = {\mnras},
         year = 2018,
        month = apr,
       volume = {475},
       number = {3},
        pages = {4066-4087},
          doi = {10.1093/mnras/sty084},
archivePrefix = {arXiv},
       eprint = {1801.05427},
 primaryClass = {astro-ph.GA},
       adsurl = {https://ui.adsabs.harvard.edu/abs/2018MNRAS.475.4066V}
}

@ARTICLE{ErraniNavarro2021,
       author = {{Errani}, Rapha{\"e}l and {Navarro}, Julio F.},
        title = "{The asymptotic tidal remnants of cold dark matter subhaloes}",
      journal = {\mnras},
         year = 2021,
        month = jul,
       volume = {505},
       number = {1},
        pages = {18-32},
          doi = {10.1093/mnras/stab1215},
archivePrefix = {arXiv},
       eprint = {2011.07077},
 primaryClass = {astro-ph.GA},
       adsurl = {https://ui.adsabs.harvard.edu/abs/2021MNRAS.505...18E}
}

@ARTICLE{PontzenGovernato2012,
       author = {{Pontzen}, Andrew and {Governato}, Fabio},
        title = "{How supernova feedback turns dark matter cusps into cores}",
      journal = {\mnras},
         year = 2012,
        month = apr,
       volume = {421},
       number = {4},
        pages = {3464-3471},
          doi = {10.1111/j.1365-2966.2012.20571.x},
archivePrefix = {arXiv},
       eprint = {1106.0499},
 primaryClass = {astro-ph.CO},
       adsurl = {https://ui.adsabs.harvard.edu/abs/2012MNRAS.421.3464P}
}

@ARTICLE{Read2016,
       author = {{Read}, J.~I. and {Agertz}, O. and {Collins}, M.~L.~M.},
        title = "{Dark matter cores all the way down}",
      journal = {\mnras},
         year = 2016,
        month = jul,
       volume = {459},
       number = {3},
        pages = {2573-2590},
          doi = {10.1093/mnras/stw713},
archivePrefix = {arXiv},
       eprint = {1508.04143},
 primaryClass = {astro-ph.GA},
       adsurl = {https://ui.adsabs.harvard.edu/abs/2016MNRAS.459.2573R}
}

@ARTICLE{Ferrero2021,
       author = {{Ferrero}, Ismael and {Navarro}, Julio F. and {Abadi}, Mario G. and {Benavides}, Jos{\'e} A. and {Mast}, Dami{\'a}n},
        title = "{A unified scenario for the origin of spiral and elliptical galaxy structural scaling laws}",
      journal = {\aap},
         year = 2021,
        month = apr,
       volume = {648},
          eid = {A124},
        pages = {A124},
          doi = {10.1051/0004-6361/202039839},
archivePrefix = {arXiv},
       eprint = {2009.03916},
 primaryClass = {astro-ph.GA},
       adsurl = {https://ui.adsabs.harvard.edu/abs/2021A&A...648A.124F}
}

@ARTICLE{vanDokkum2022,
       author = {{van Dokkum}, Pieter and {Shen}, Zili and {Keim}, Michael A. and {Trujillo-Gomez}, Sebastian and {Danieli}, Shany and {Dutta Chowdhury}, Dhruba and {Abraham}, Roberto and {Conroy}, Charlie and {Kruijssen}, J.~M. Diederik and {Nagai}, Daisuke and {Romanowsky}, Aaron},
        title = "{A trail of dark-matter-free galaxies from a bullet-dwarf collision}",
      journal = {\nat},
         year = 2022,
        month = may,
       volume = {605},
       number = {7910},
        pages = {435-439},
          doi = {10.1038/s41586-022-04665-6},
archivePrefix = {arXiv},
       eprint = {2205.08552},
 primaryClass = {astro-ph.GA},
       adsurl = {https://ui.adsabs.harvard.edu/abs/2022Natur.605..435V}
}

@ARTICLE{vanDokkum2019,
       author = {{van Dokkum}, Pieter and {Danieli}, Shany and {Abraham}, Roberto and {Conroy}, Charlie and {Romanowsky}, Aaron J.},
        title = "{A Second Galaxy Missing Dark Matter in the NGC 1052 Group}",
      journal = {\apjl},
         year = 2019,
        month = mar,
       volume = {874},
       number = {1},
          eid = {L5},
        pages = {L5},
          doi = {10.3847/2041-8213/ab0d92},
archivePrefix = {arXiv},
       eprint = {1901.05973},
 primaryClass = {astro-ph.GA},
       adsurl = {https://ui.adsabs.harvard.edu/abs/2019ApJ...874L...5V}
}

@ARTICLE{vanDokkum2018,
       author = {{van Dokkum}, Pieter and {Danieli}, Shany and {Cohen}, Yotam and {Merritt}, Allison and {Romanowsky}, Aaron J. and {Abraham}, Roberto and {Brodie}, Jean and {Conroy}, Charlie and {Lokhorst}, Deborah and {Mowla}, Lamiya and {O'Sullivan}, Ewan and {Zhang}, Jielai},
        title = "{A galaxy lacking dark matter}",
      journal = {\nat},
         year = 2018,
        month = mar,
       volume = {555},
       number = {7698},
        pages = {629-632},
          doi = {10.1038/nature25767},
archivePrefix = {arXiv},
       eprint = {1803.10237},
 primaryClass = {astro-ph.GA},
       adsurl = {https://ui.adsabs.harvard.edu/abs/2018Natur.555..629V}
}

@ARTICLE{Navarro1996a,
       author = {{Navarro}, Julio F. and {Frenk}, Carlos S. and {White}, Simon D.~M.},
        title = "{The Structure of Cold Dark Matter Halos}",
      journal = {\apj},
         year = 1996,
        month = may,
       volume = {462},
        pages = {563},
          doi = {10.1086/177173},
archivePrefix = {arXiv},
       eprint = {astro-ph/9508025},
 primaryClass = {astro-ph},
       adsurl = {https://ui.adsabs.harvard.edu/abs/1996ApJ...462..563N}
}

@ARTICLE{Navarro1996b,
       author = {{Navarro}, Julio F. and {Eke}, Vincent R. and {Frenk}, Carlos S.},
        title = "{The cores of dwarf galaxy haloes}",
      journal = {\mnras},
         year = 1996,
        month = dec,
       volume = {283},
       number = {3},
        pages = {L72-L78},
          doi = {10.1093/mnras/283.3.L72},
archivePrefix = {arXiv},
       eprint = {astro-ph/9610187},
 primaryClass = {astro-ph},
       adsurl = {https://ui.adsabs.harvard.edu/abs/1996MNRAS.283L..72N}
}

@ARTICLE{Mercado2025,
       author = {{Mercado}, Francisco J. and {Moreno}, Jorge and {Feldmann}, Robert and {Zeender}, Marckie and {Benavides}, Jos{\'e} A. and {Piotrowska}, Joanna M. and {Klein}, Courtney and {Wheeler}, Coral and {Necib}, Lina and {Bullock}, James S. and {Hopkins}, Philip F.},
        title = "{Effects of Galactic Environment on Size and Dark Matter Content in Low-mass Galaxies}",
      journal = {\apj},
         year = 2025,
        month = apr,
       volume = {983},
       number = {2},
          eid = {93},
        pages = {93},
          doi = {10.3847/1538-4357/adbf07},
archivePrefix = {arXiv},
       eprint = {2501.04084},
 primaryClass = {astro-ph.GA},
       adsurl = {https://ui.adsabs.harvard.edu/abs/2025ApJ...983...93M}
}

@ARTICLE{Sharda2026,
       author = {{Sharda}, Piyush and {Schaye}, Joop and {McGibbon}, Robert J. and {Ben{\'\i}tez-Llambay}, Alejandro and {Chaikin}, Evgenii and {Frenk}, Carlos S. and {Hodge}, Jacqueline and {Hu{\v{s}}ko}, Filip and {Ploeckinger}, Sylvia and {Richings}, Alexander J. and {Schaller}, Matthieu},
        title = "{The evolution of the galaxy gas-phase mass-metallicity relation from $z=15$ to $z=0$ in the COLIBRE cosmological simulations}",
      journal = {arXiv e-prints},
         year = 2026,
        month = jun,
          eid = {arXiv:2606.25995},
        pages = {arXiv:2606.25995},
          doi = {10.48550/arXiv.2606.25995},
archivePrefix = {arXiv},
       eprint = {2606.25995},
 primaryClass = {astro-ph.GA},
       adsurl = {https://ui.adsabs.harvard.edu/abs/2026arXiv260625995S}
}

@ARTICLE{BL2019,
       author = {{Ben{\'\i}tez-Llambay}, Alejandro and {Frenk}, Carlos S. and {Ludlow}, Aaron D. and {Navarro}, Julio F.},
        title = "{Baryon-induced dark matter cores in the EAGLE simulations}",
      journal = {\mnras},
         year = 2019,
        month = sep,
       volume = {488},
       number = {2},
        pages = {2387-2404},
          doi = {10.1093/mnras/stz1890},
archivePrefix = {arXiv},
       eprint = {1810.04186},
 primaryClass = {astro-ph.GA},
       adsurl = {https://ui.adsabs.harvard.edu/abs/2019MNRAS.488.2387B}
}

@ARTICLE{Lagos2026,
       author = {{Lagos}, Claudia del P. and {Schaye}, Joop and {Schaller}, Matthieu and {Obreschkow}, Danail and {Bah{\'e}}, Yannick M. and {Ben{\'\i}tez-Llambay}, Alejandro and {Chaikin}, Evgenii and {Correa}, Camila and {Davis}, Timothy A. and {Frenk}, Carlos S. and {Hu{\v{s}}ko}, Filip and {Kaasinen}, Melanie and {McGibbon}, Robert J. and {Oman}, Kyle and {Ploeckinger}, Sylvia and {Richings}, Alexander J. and {Trayford}, James W. and {Wang}, Jing and {Wright}, Ruby J.},
        title = "{Kennicutt─Schmidt relation of galaxies over 13 billion years in the COLIBRE hydrodynamical simulations}",
      journal = {\mnras},
         year = 2026,
        month = jun,
       volume = {549},
       number = {2},
          eid = {stag947},
        pages = {stag947},
          doi = {10.1093/mnras/stag947},
archivePrefix = {arXiv},
       eprint = {2512.11309},
 primaryClass = {astro-ph.GA},
       adsurl = {https://ui.adsabs.harvard.edu/abs/2026MNRAS.549ag947L}
}

@ARTICLE{Revaz2018,
       author = {{Revaz}, Yves and {Jablonka}, Pascale},
        title = "{Pushing back the limits: detailed properties of dwarf galaxies in a {\ensuremath{\Lambda}}CDM universe}",
      journal = {\aap},
         year = 2018,
        month = aug,
       volume = {616},
          eid = {A96},
        pages = {A96},
          doi = {10.1051/0004-6361/201832669},
archivePrefix = {arXiv},
       eprint = {1801.06222},
 primaryClass = {astro-ph.GA},
       adsurl = {https://ui.adsabs.harvard.edu/abs/2018A&A...616A..96R}
}

@misc{deAlmeida2026,
      title={Satellite compaction pathways: environmental drivers shaping dwarf galaxy corpulence in the TNG50 simulation}, 
      author={Abhner P. De Almeida and Gary A. Mamon and Gastão B. Lima Neto},
      year={2026},
      eprint={2606.09817},
      archivePrefix={arXiv},
      primaryClass={astro-ph.GA},
      url={https://arxiv.org/abs/2606.09817}, 
}

@ARTICLE{Nelson2018,
       author = {{Nelson}, Dylan and {Pillepich}, Annalisa and {Springel}, Volker and {Weinberger}, Rainer and {Hernquist}, Lars and {Pakmor}, R{\"u}diger and {Genel}, Shy and {Torrey}, Paul and {Vogelsberger}, Mark and {Kauffmann}, Guinevere and {Marinacci}, Federico and {Naiman}, Jill},
        title = "{First results from the IllustrisTNG simulations: the galaxy colour bimodality}",
      journal = {\mnras},
         year = 2018,
        month = mar,
       volume = {475},
       number = {1},
        pages = {624-647},
          doi = {10.1093/mnras/stx3040},
archivePrefix = {arXiv},
       eprint = {1707.03395},
 primaryClass = {astro-ph.GA},
       adsurl = {https://ui.adsabs.harvard.edu/abs/2018MNRAS.475..624N}
}

\begin{appendix}

\section{Numerical convergence} \label{AppNumConv}

We briefly discuss the convergence of the Stellar Mass-Size Relation (SMSR) for central and satellite galaxies within the COLIBRE framework \citep[for a detailed discussion on numerical convergence across the suite, see][]{Schaye2025,Ludlow2026,Forouhar2026}. To ensure a consistent comparison, we utilize different particle resolutions within the same cosmological volume, i.e. we compare the L200m6 run against L200m7, and to reach higher resolution we compare the significantly smaller (a factor $8^3 = 512$) volume L025m5 against L025m6 and L025m7 (where ``m5'', ``m6'' and ``m7'' indicate that DM and stars particles have masses of $\sim$$10^5$, $\sim$$10^6$ and $\sim$$10^7 \, {\rm M_{\odot}}$). Runs with the same cosmological volume have the same initial conditions, but simulate fewer, more massive particles than the fiducial runs. The L025 run contains far fewer galaxies with $M_* > 10^8 \, {\rm M_{\odot}}$ ($\sim$700 total centrals and satellites, compared to $\sim$300,000 in the $200^3 \, {\rm Mpc^3}$ boxes) and lacks massive groups.

\begin{figure}
    \centering
    \includegraphics[width=\columnwidth]{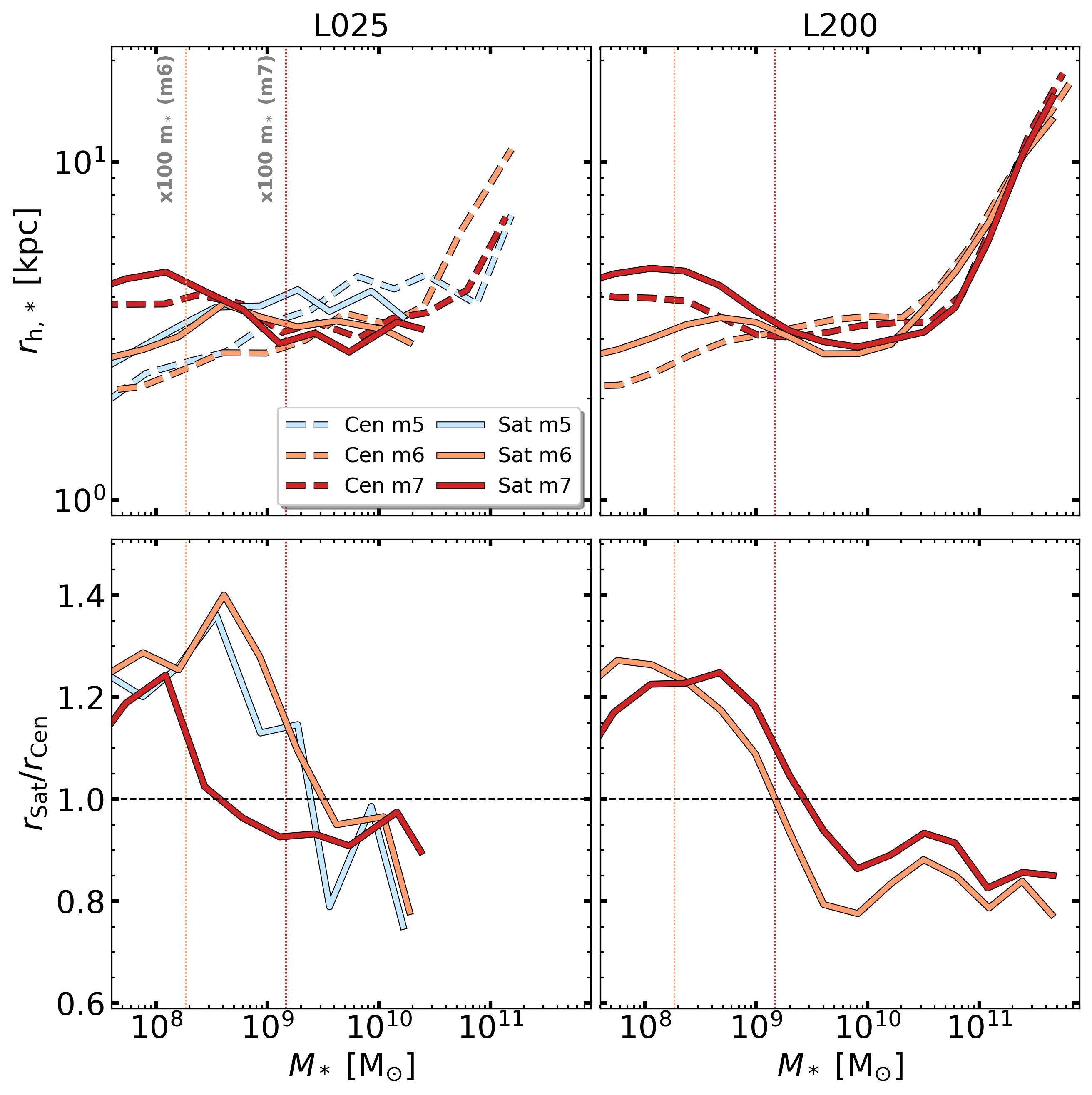}
    
    \caption{Convergence test within the COLIBRE suite for two different simulated volumes: L025 ($25^3 \, {\rm cMpc^3}$, left panels) and L200 ($200^3 \, {\rm cMpc^3}$, right panels). Top panels: median trends of stellar half-mass radius ($r_{h,*}$, 3D) as a function of stellar mass ($M_{*}$) for $z=0$ centrals (dashed lines) and satellites (solid lines) from different resolutions available: m5 (light blue), m6 (orange) and m7 (red). Bottom panels: ratio between the satellites' trend and that of centrals. We indicate with vertical dotted lines the 100 stellar particle limit for m7 (red, m$_* = 1.47 \times 10^9 \, {\rm M_{\odot}}$) and m6 (orange, m$_* = 1.84 \times 10^8 \, {\rm M_{\odot}}$), m5's falls out of the limits of the figure.}
    \label{Fig1App}
\end{figure}

In Fig. \ref{Fig1App} we show the SMSR for centrals (dashed lines) and satellites (solid lines) for the two different simulated volumes: L025 ($25^3 \, {\rm Mpc^3}$, left panels) and L200 ($200^3 \, {\rm Mpc^3}$, right panels). Each particle mass run is highlighted with a different colour: m5 light blue, m6 orange, and m7 red. We also show the stellar mass limit corresponding to 100 particles for each resolution with vertical lines (m5's falls below the plotted range).

We find good agreement across all runs until the stellar particle count drops below 100, most noticeable by the size increase of central galaxies from the m7 runs. Notably, dwarf satellites ($\log(M_*/{\rm M_\odot}) < 9$) exhibit larger sizes than centrals with similar mass across all runs, and the opposite is true in the bright regime ($9 < \log(M_*/{\rm M_\odot}) < 10.5$). However, satellites from the L025m5 run with $9.5 < \log(M_*/{\rm M_\odot})$ appear systematically larger; this is likely due to the absence of cluster environments in the smaller volume, allowing these galaxies to grow for longer in lower-density regions. These difference are more evident in the bottom panels, where we show the ratio between the satellite SMSR and that of centrals.

The runs L200m6 and L200m7, which sample more (massive) galaxies, converge in the stellar mass range $9.3 \lesssim \log(M_*/{\rm M_\odot})$, where m7's galaxies have more than 100 stellar particles. For lower masses ($8 \lesssim \log(M_*/{\rm M_\odot}) < 9.3$), the L025m6 run closely follows the median trends of the higher resolution L025m5. Therefore, L200m6 appears as the optimal trade-off between particle resolution and volume sample for this study.

\section{Tidal evolutionary tracks}
\label{AppTET}

\begin{figure}
    \centering
    \includegraphics[width=\columnwidth]{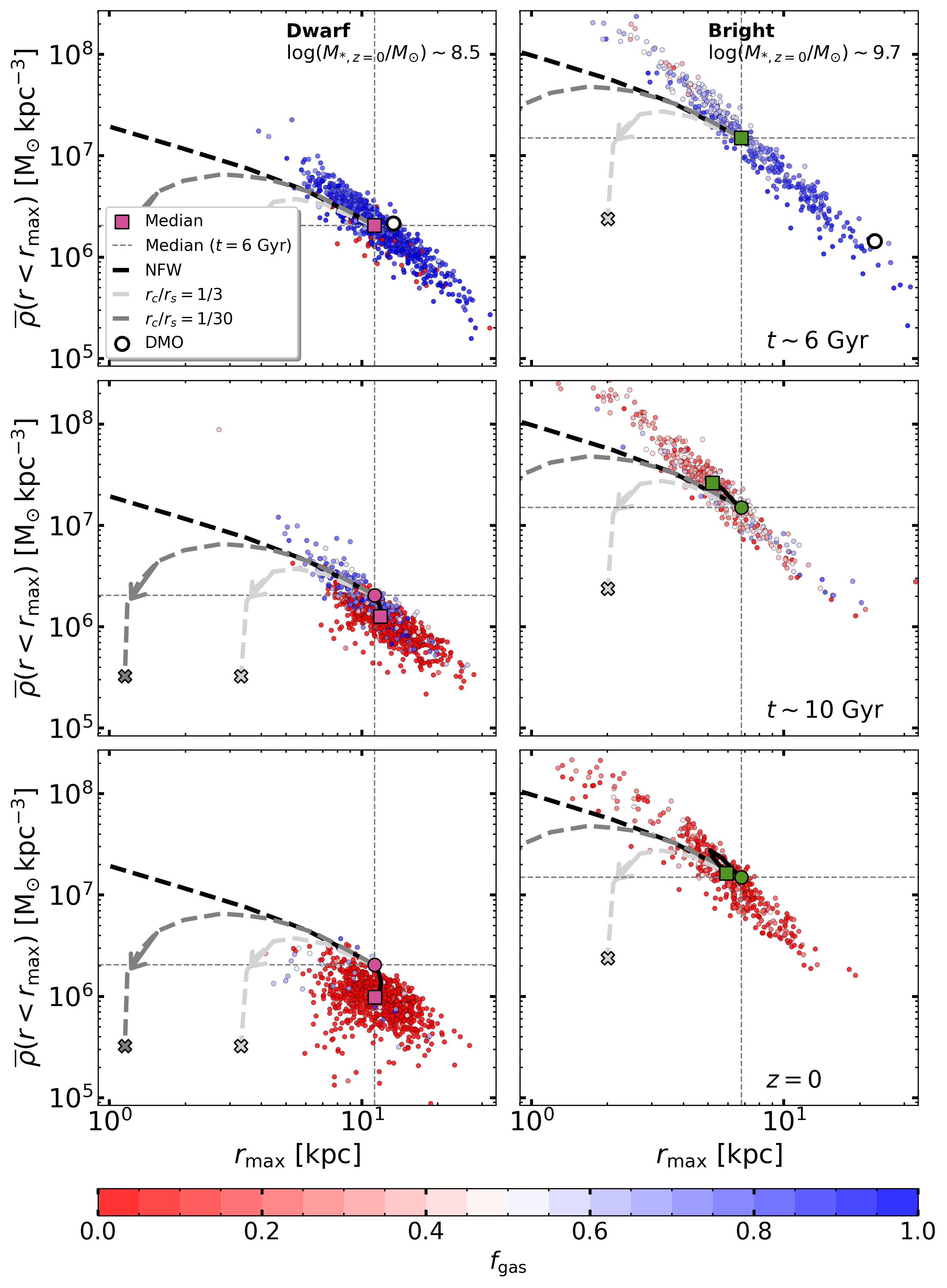}
    
    \caption{Tidal Evolutionary Tracks (TET) of the dwarf (left) and bright (right) satellite subsamples (``pink'' and ``dark green'' bins, respectively, in Fig. \ref{SMSR_tracks_subsamples}). We show the mean density within the maximum circular velocity radius $\bar{\rho} (r < r_{\rm max}) = 3/4 \pi G \, (V_{\rm max}/r_{\rm max})^2$, with $V_{\rm max}$ the maximum circular velocity, as a function of the radius of maximum circular velocity $r_{\rm max}$. Each galaxy is shown as a circle coloured by the gas fraction $f_{\rm gas}$ at three cosmic times. From top to bottom, we show: i) $t=6$ Gyr, prior to infall; ii) $t=10$ Gyr, when some galaxies already have lost most of their gas (dwarfs lose baryons, while bright satellites transform it into stars); and iii) $z=0$. The median evolutionary tracks are shown as solid black lines, from $t=6$ Gyr (coloured circles) to $z=0$ (coloured squares). For comparison, we show the TET of a pure NFW distribution (black dashed line), and that of a cored-NFW profile with core size $r_{\rm c}/r_{\rm s} = 1/30$ (grey dashed line) and $r_{\rm c}/r_{\rm s} = 1/3$ (light grey dashed line), as presented in \citet[][]{Errani23} but re-normalised to match the median values at $t = 6$ Gyr and followed until disruption shown with crosses. For comparison, the median values of DMO subhaloes with similar virial mass at $t=6$ Gyr are shown with white circles in the top panel. Dwarfs evolve as cores in ongoing disruption, while bright satellites evolve as cuspier mass distributions due to higher central stellar mass fractions.}
    \label{TET_subsamples}
\end{figure}

The expansion of dwarf satellites during ongoing tidal disruption is compelling, but it is important to validate it with a more detailed analysis. We attempt to do this here by using the Tidal Evolutionary Track (TET) formalism from \cite{Errani23}. This framework characterizes mass distributions via their circular velocity profiles, specifically by the maximum circular velocity, $V_{\rm max}$, the radius at which this velocity is attained, $r_{\rm max}$, and the period of a circular orbit at that radius, $T_{\rm max} = 2 \, \pi \, r_{\rm max}/V_{\rm max}$. Because the orbital period is inversely proportional to the square root of the density $(T \propto \rho^{-1/2})$, we instead adopt the mean density of the enclosed mass, defined as:
\begin{equation}
\label{eq_bar_rho}
    \bar{\rho} (r < r_{\rm max}) = \frac{3 \, V^2_{\rm max}}{4 \pi G \, r_{\rm max}^2} 
\end{equation}

Fig. \ref{TET_subsamples} illustrates the evolution of the mean internal density $\bar{\rho} (r < r_{\rm max})$ as function of $r_{\rm max}$ for our two illustrative satellite subsamples: dwarf (left panels) and bright (right panels) satellites. We show three cosmic times, from top to bottom: $t = 6$ Gyr (pre-infall), $t = 10$ Gyr and $z=0$. Each galaxy is represented by a circle coloured by its instantaneous gas fraction $f_{\rm gas} = M_{\rm gas}/M_{\rm baryon}$. Dashed lines indicate the theoretical tidal evolution of different mass distributions embedded in a static, isothermal host potential, as reported by \cite{Errani23}. A cuspy mass distribution such as an NFW (shown in black), becomes denser as it loses its least-bound outer mass, moving upwards and to the left in the ($r_{\rm max}$, $\bar{\rho}$) plane. In contrast, a ``cored-NFW'' profile \citep{Penarrubia2012} eventually undergoes tidal disruption, becoming less dense as it loses mass. Larger core sizes $r_{\rm c}$ (expressed in units of the scale radius $r_{\rm s}$) mean that this disruption begins at earlier stages of mass loss; profiles with $r_{\rm c}/r_{\rm s} = 1/30$ (grey) and with $r_{\rm c}/r_{\rm s} = 1/3$ (light grey) moving downwards toward lower densities at different values of $r_{\rm max}$. Literature values were transformed into $\bar{\rho}$ using Eq. \ref{eq_bar_rho} and renormalised to match the median values of our subsamples at $t = 6$ Gyr.

For comparison, we also show median values of central subhaloes from the DM-Only run of L200m6 (``DMO'') that have similar virial mass (white circles in the top panel). These indicate that dwarf galaxies, prior to infall, have similar properties as DMO subhaloes and thus that baryons do not significantly affect the total mass distribution. On the other hand, bright satellites show higher densities than those of DMO subhaloes with similar mass due to the presence of a non-negligible stellar component, as shown in Sec. \ref{subsec:SMSR_TemporalEvolution}.

Dwarf galaxies do not follow the TET of cuspy mass distributions. Prior to infall, these systems are characterised by high gas fractions $f_{\rm gas}$ (bluer colours); however, as they undergo gas stripping\footnote{Dwarf galaxies only grow slightly in stellar mass, suggesting that the gas fraction decrease is not due to star formation but rather due to large gas-mass losses by stripping.} (lower $f_{\rm gas}$, redder colours), they migrate vertically downwards in the ($r_{\rm max}$, $\bar{\rho}$) plane. This progressive density decrease at near-constant $r_{\rm max}$ is a characteristic behaviour of tidal disruption in systems with prominent cores.

Bright satellites, in contrast, exhibit only mild evolution in this plane; they neither behave as cored mass distributions nor follow the TET characteristic of cuspy profiles. These galaxies retain most of their baryons at the innermost regions and continue to form stars post-infall, resulting in a median stellar mass growth of $\sim$0.5 dex (as shown in Fig. \ref{SMSR_tracks_subsamples}). Consequently, the presence of stars maintain high, stable densities near the galactic centres, hindering a direct comparison with gravity-only N-body simulations of cuspier profiles \citep[see e.g.][]{Governato2012,DiCintio2014,Lazar2020}. Therefore, the TETs of DMO subhaloes cannot provide quantitative predictions for satellites in the stellar mass range $8 < \log(M_\star/\mathrm{M}_\odot) < 10.5$, as the influence of baryons in the innermost regions significantly alters the evolution of these systems.

\end{appendix}
\end{document}